\documentclass[twocolumn]{aastex63}

\usepackage{graphicx, epsfig}
\usepackage{color}
\usepackage{mathrsfs}
\usepackage{bm}
\usepackage{amsmath,amssymb,yhmath}
\usepackage[caption=false]{subfig}
\usepackage{hyperref}
\usepackage[normalem]{ulem}
\usepackage{mathtools}
\usepackage{float}
\newcommand{\be}{\begin{equation}}
\newcommand{\ee}{\end{equation}}
\newcommand{\beq}{\begin{eqnarray}}
\newcommand{\eeq}{\end{eqnarray}}

\def\barnue{\mathrel{{\bar \nu}_e}}

\def \sun	 	{$_{\odot}$}

\def\t13{\mathrel{{\theta_{13}}}}
\def\y12{\mathrel{{\tan^2 \theta_{12}}}}
\def\c2{\mathrel{{\chi^2 }}}

\def\msun{\mathrel{{M_{\odot} }}}

\def \sun	 	{$_{\odot}$}

\newcommand{\n}{neutrino}
\newcommand{\ns}{neutrinos}
\newcommand{\ps}{presupernova}
\newcommand{\sn}{supernova}

\newcommand{\ls}{LS}
\newcommand{\lli}{LS-Li}

\shorttitle{Pre-SN Localization}
\shortauthors{Mukhopadhyay et al.}

\begin{document}

\title{Presupernova neutrinos: directional sensitivity and prospects for progenitor identification}

\author[0000-0002-2109-5315]{Mainak Mukhopadhyay}
\affiliation{Department of Physics, Arizona State University, Tempe, AZ 85287, USA.}
\author[0000-0002-9253-1663]{Cecilia Lunardini}
\affiliation{Department of Physics, Arizona State University, Tempe, AZ 85287, USA.}
\author[0000-0002-0474-159X]{F.X.~Timmes}
\affiliation{School of Earth and Space Exploration, Arizona State University, Tempe, AZ 85287, USA.}
\affiliation{Joint Institute for Nuclear Astrophysics - Center for the Evolution of the Elements, USA.}
\author[0000-0001-8689-4495]{Kai Zuber}
\affiliation{Institute for Nuclear and Particle Physics, TU Dresden, 01069 Dresden, Germany.}

\begin{abstract}
We explore the potential of current and future liquid scintillator
neutrino detectors of $\mathcal O (10)$ kt mass to localize a
pre-supernova neutrino signal in the sky. In the hours preceding the
core collapse of a nearby star (at distance $D \lesssim 1$ kpc), tens
to hundreds of inverse beta decay events will be recorded, and their
reconstructed topology in the detector can be used to estimate the
direction to the star. Although the directionality of inverse beta
decay is weak ($\sim$\,8\% forward-backward asymmetry for currently
available liquid scintillators), we find that for a fiducial signal of
$200$ events (which is realistic for Betelgeuse), a positional error
of $\sim$\,60$^\circ$ can be achieved, resulting in the possibility to
narrow the list of potential stellar candidates to less than ten,
typically. For a configuration with improved forward-backward
asymmetry ($\sim$\,40\%, as expected for a lithium-loaded liquid
scintillator), the angular sensitivity improves to $\sim$\,15$^\circ$,
and -- when a distance upper limit is obtained from the overall event
rate -- it is in principle possible to uniquely identify the
progenitor star.  Any localization information accompanying an early
supernova alert will be useful to multi-messenger observations
and to particle physics tests using collapsing stars.
\end{abstract}

\keywords{
Neutrino astronomy (1100),
Neutrino telescopes (1105),
Supernova neutrinos (1666),
High energy astrophysics (739)
}

\section{Introduction}
\label{intro}

Over the next decade, neutrino astronomy will probe the rich
astrophysics of neutrino production in the sky. In addition to neutrinos from
the Sun \citep{borexino-collaboration_2018_aa},
core-collapse supernova bursts \citep[e.g., SN 1987A,][]{hirata_1987_aa,hirata_1988_aa, bionta_1987_aa, alekseev_1987_aa},
and relativistic jets 
\citep[e.g., blazar TXS 0506+056,][]{icecube-collaboration_2018_aa,icecube-collaboration_2018_ab}, 
technological improvements in detector masses, energy resolution and
background abatement will allow to observe \emph{new} signals from
different stages of the lifecycle of stars, in particular
\ps\ \ns\ \citep{Odrzywolek:2003vn}, the diffuse supernova neutrino background
\citep{NYAS:NYAS319,Krauss:1983zn}, and \ns\ from matter-rich
binary mergers \citep{Kyutoku:2017wnb,Lin:2019piz}. Ultimately, the goal will be to
test neutrino production across the entire Hertzsprung-Russell diagram
\citep{farag_2020_aa}.

Presupernova \ns\ are the \ns\ of $\sim$ 0.1 - 5 MeV energy that
accompany, with increasing luminosity, the last stages of nuclear
burning of a massive star in the days leading to its core collapse and
final explosion as a supernova, or implosion into a black hole 
(a ``failed'' supernova). These \ns\ are produced by thermal processes --
mainly pair-production -- that depend on the ambient thermodynamic
conditions \citep{fowler_1964_aa, beaudet_1967_aa,schinder_1987_aa,itoh_1996_aa} -- and by weak
reactions -- mainly electron/positron captures and nuclear decays --
that have a stronger dependence on the isotopic composition
\citep{fuller_1980_aa,fuller_1982_aa,fuller_1982_ab,fuller_1985_aa,
langanke_2000_aa,langanke_2014_aa,misch_2018_aa}, and thus on the
network of nuclear reactions that take place in the stellar interior.

Building on early calculations \citep{Odrzywolek:2003vn, Odrzywolek:2004em,Kutschera:2009ff, odrzywolek_2009_aa}, recent numerical simulations
with state-of-the-art treatment of the nuclear processes \citep{kato_2015_aa,yoshida_2016_aa,patton_2017_aa,patton_2017_ab,kato_2017_aa,Guo:2019orq} 
have shown that the \ps\ \n\ flux increases dramatically, both in
luminosity and in average energy, in the hours prior to the collapse,
and it becomes potentially detectable when silicon burning is ignited
in the core of the star.  In particular, for stars within $\sim$\,1~kpc
of Earth like Betelgeuse, \ps\ \ns\ will be detected at multi-kiloton
\n\ detectors like the  current KamLAND (see \cite{Araki:2004mb} for a
dedicated study), Borexino \citep{borexino-collaboration_2018_aa},
SNO+ \citep{Andringa:2015tza}, Daya Bay \citep{Guo:2007ug} and
SuperKamiokande \citep{simpson_2019_aa}, and the upcoming
HyperKamiokande \citep{abe_2016_aa}, DUNE \citep{Acciarri:2016ooe} and
JUNO \citep{an_2016_aa,li_2014_aa,brugiere_2017_aa}. Next generation
dark matter detectors like XENON \citep{newstead_2019_aa}, DARWIN
\citep{aalbers_2016_aa}, and ARGO \citep{aalseth_2018_aa} will also
observe a significant signal \citep{Raj:2019wpy}.  Therefore, \ps\ \ns\ are a
prime target for the SuperNova Early Warning System network
\citep[SNEWS,][]{antonioli_2004_aa} -- which does or will include the
\n\ experiments mentioned above -- and its multi-messenger era
successor SNEWS 2.0, whose mission is to provide early alerts
to the astronomy and gravitational wave communities, and to the
scientific community at large as well. The observation of presupernova
neutrinos from an impending core-collapse supernova will: (i) allow
numerous tests of stellar and neutrino physics, including tests of
exotic physics that may require pointing to the collapsing star 
(e.g. axion searches, see \cite{Raffelt:2011ft}); and (ii) enable a very early alert of
the collapse and supernova, thus extending  -- perhaps crucially,
especially for envelope-free stellar progenitors that tend to explode
shortly after collapse -- the time frame available to coordinate
multi-messenger observations.

In this paper, we explore \ps\ \ns\ as early alerts. In particular, we
focus on the question of localization: can a signal of presupernova
neutrinos provide useful positional information? Can it identify the
progenitor star?  From a recent exploratory study \citep{li_2020_aa},
we know that the best potential for localization is offered by inverse
beta decay events at large (${\mathcal O}(10)$ kt mass) liquid
scintillator detectors, where, for optimistic \ps\ flux predictions
and a star like Betelgeuse (distance of 0.2 kpc), a signal can be
discovered days before the collapse, and the direction to the
progenitor can be determined with a $\sim 80^{\circ}$ error.  

This article is the first dedicated study on the localization question for \ps\ neutrinos. 
Using a state-of-the-art numerical model for the \n\ emission, we examine a number of questions that were not previously discussed, having to do with the diverse stellar population of nearby stars (including red and blue supergiants, of masses between $\sim$\,10 and $\sim$\,30 times the mass of the Sun, and clustered in certain regions of the sky) and with the rich possibilities of improving the directionality of the liquid scintillator technology in the future.

In Section~\ref{sec:rates} we discuss \ps\ neutrino event rates and nearby candidates.
In Section~\ref{sec:method} we present our main results for the angular sensitivity.
In Section~\ref{sec:progenid} we discuss progenitor identification, and
in Section~\ref{sec:discussion} we summarize our results.
In Appendix~\ref{appendixA} we detail the distance and mass estimates of nearby \ps\ candidates.

\begin{figure*}[!htb]
\centering
\includegraphics[width=0.48\textwidth]{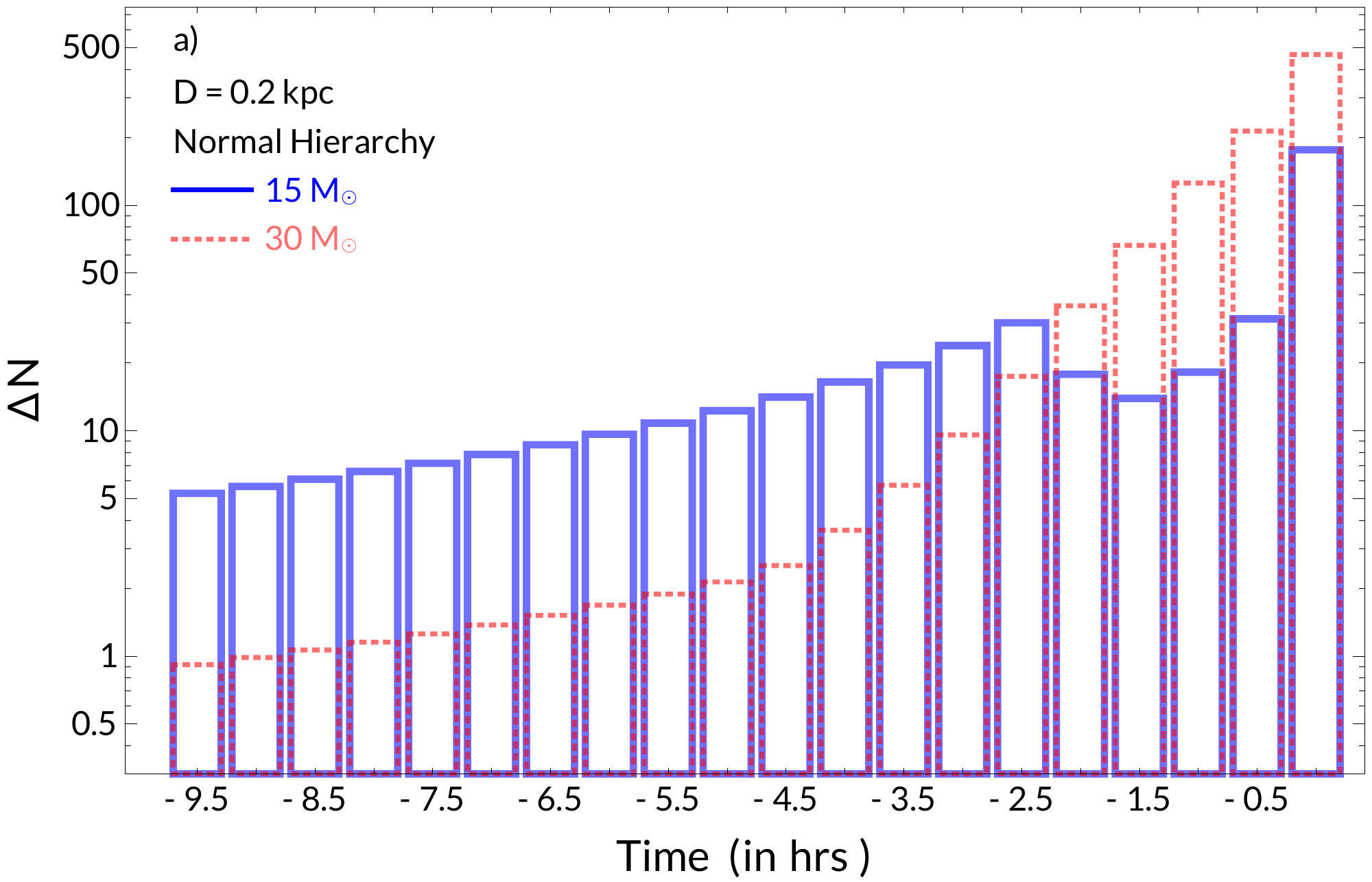} 
\includegraphics[width=0.48\textwidth]{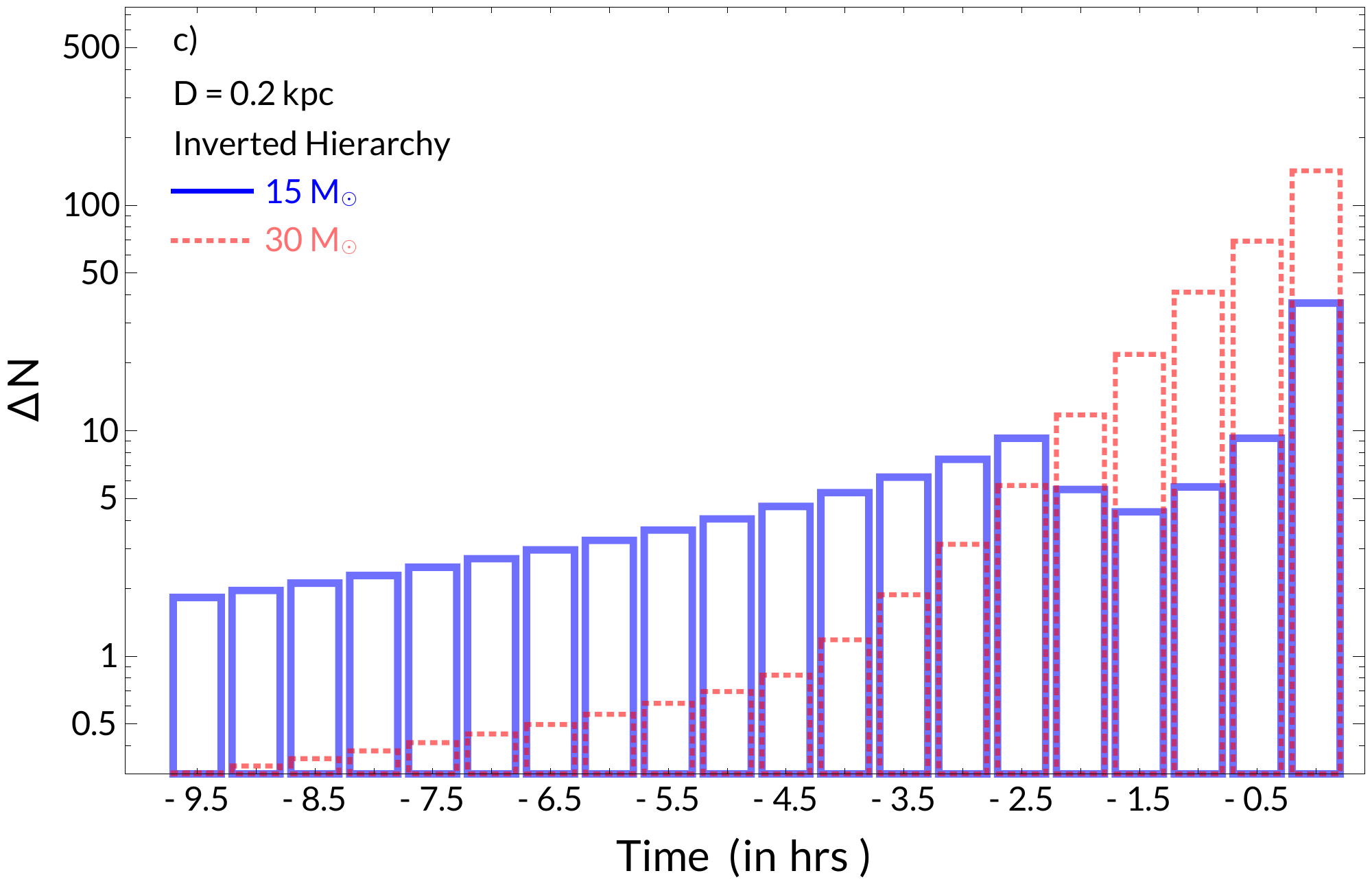} 
\includegraphics[width=0.48\textwidth]{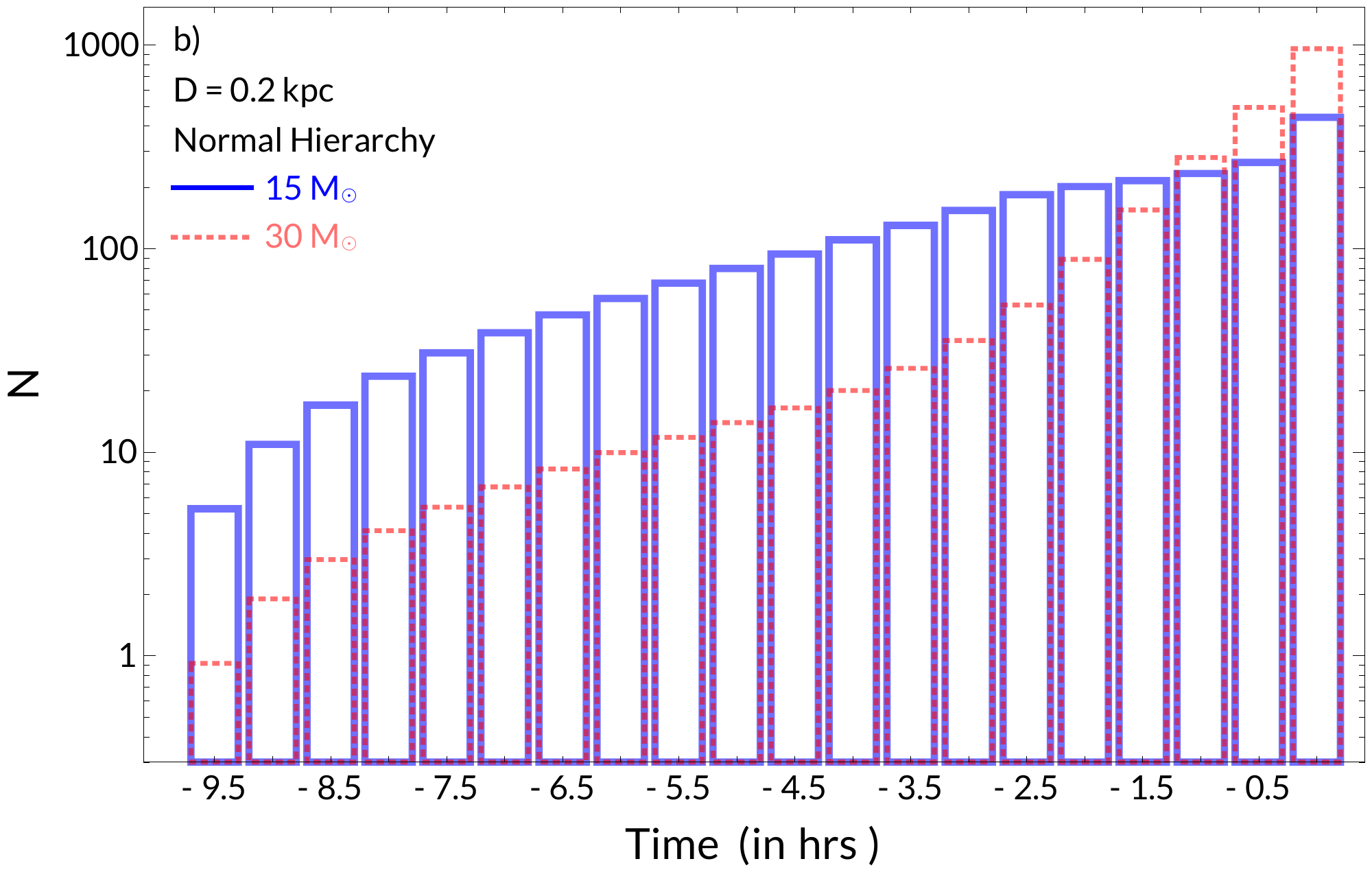} 
\includegraphics[width=0.48\textwidth]{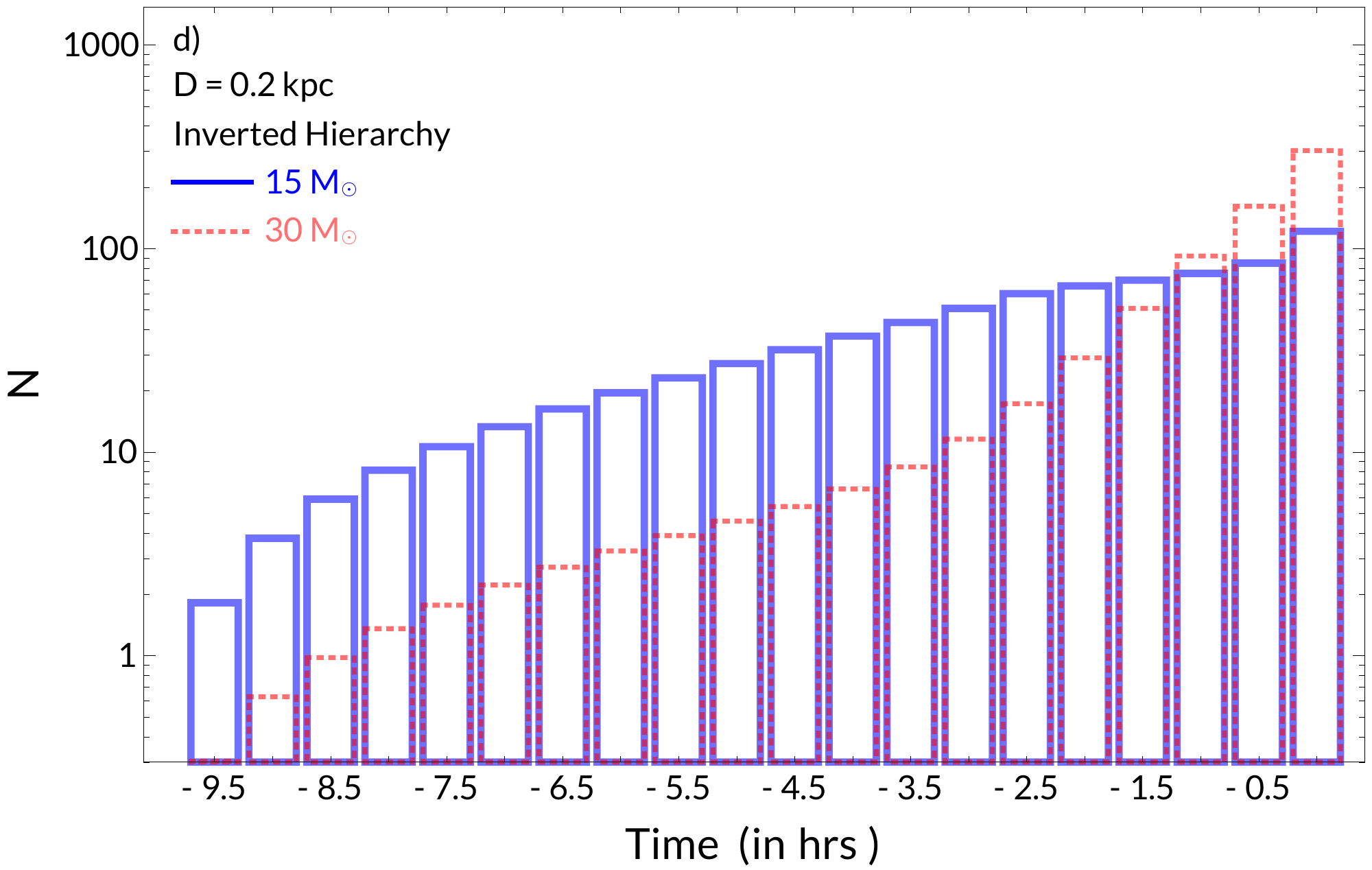}
\caption{
Top row a) and c): Number of \ps\ \n\ events at a 17 kt liquid
scintillator detector, in time bins of width $\Delta t=0.5$ hrs
as a function of time before core-collapse.
Bottom row or b) and d): Cumulative numbers of events
in half-hour increments. Shown are the cases of a ZAMS 15 $\msun$
(blue histogram) and a ZAMS 30 $\msun$ (red histogram) progenitor, at a distance
$D$=0.2~kpc, for the normal (left column)
and inverted (right column) neutrino mass hierarchy.
}
\label{fig:eventrate}
\end{figure*}

\begin{figure*}[htb]
\begin{center}
\includegraphics[width=0.95\textwidth,angle=0]{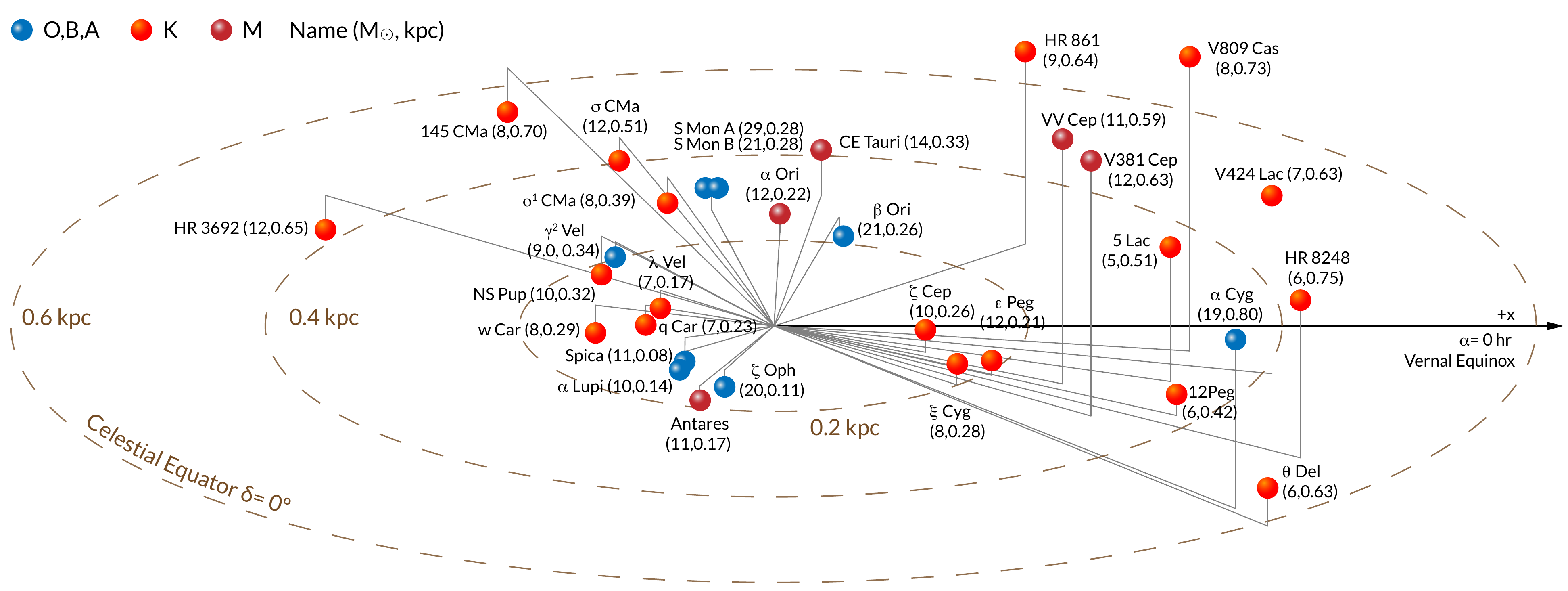}
\end{center}
\caption{
Illustration of nearby ($D\leq 1$ kpc) core collapse supernova
candidates.  Each star's spectral type, name, mass and
distance is shown in labels.  See Table \ref{tab:presncandidates} for details and references.
        }
\label{fig:nearmassive2}
\end{figure*}

\begin{figure*}[htb]
\begin{center}
\includegraphics[width=0.7\textwidth,angle=0]{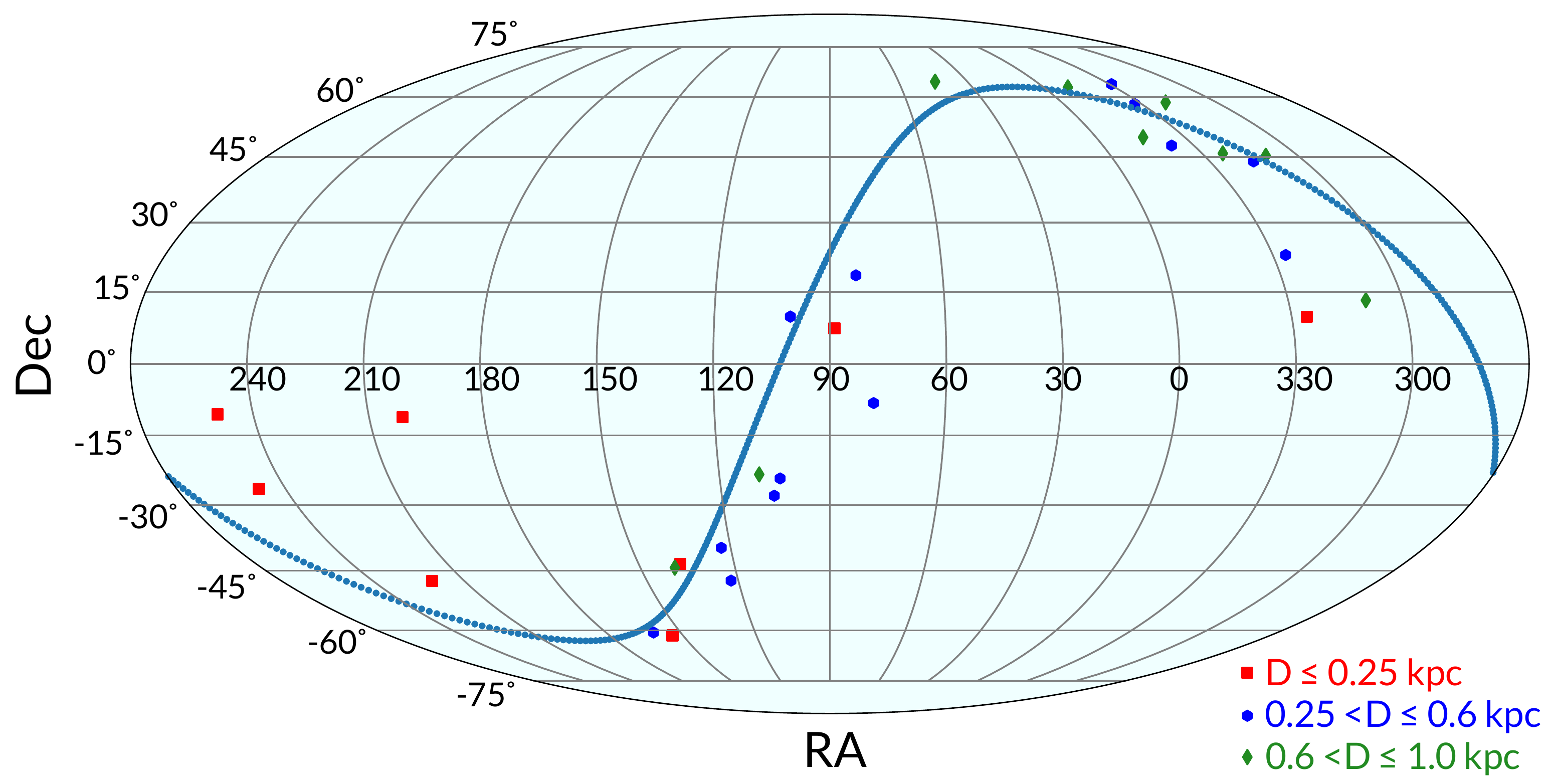}
\end{center}
\caption{
Mollweide projection of nearby ($D\leq 1$ kpc) core collapse supernova
candidates. Symbols and colors correspond to distance intervals. 
The dotted line indicates the Galactic Plane.  
The red square near the center of the map is $\alpha$ Ori, best known as Betelgeuse.
        }
\label{fig:mollweidepresn}
\end{figure*}

\section{Presupernova neutrino event rates and candidates}
\label{sec:rates}

A liquid scintillator is ideal for the detection of \ps\ \ns, through the 
inverse beta decay process (henceforth IBD, $\barnue + p \rightarrow n + e^+$)
due to its low energy threshold (1.8~MeV), and its timing, energy
resolution, and background discrimination performance.  
The expected signal from a \ps\ in neutrino detectors has been
presented in recent articles \citep[e.g.,][]{asakura_2016_aa,kato_2015_aa,yoshida_2016_aa,patton_2017_aa,kato_2017_aa,li_2020_aa}.

We consider an active detector mass of 17 kt, which is expected for JUNO, with detection efficiency of unity, and we use the
IBD event rates in \cite{patton_2017_aa,patton_kelly_m_2019_2626645}.  Figure~\ref{fig:eventrate}
shows the numbers of events and cumulative numbers of events for progenitor stars of zero
age main-sequence (ZAMS) masses of 15\,$\msun$ and 30\,$\msun$ (here $\msun=1.99\, 10^{33}$ g is the mass of the Sun) at a
distance of $D$=0.2~kpc (representative of Betelgeuse). Results are shown 
for the normal and inverted hierarchy of the \n\ mass spectrum.  
Times are negative, being relative to the time of core-collapse.

Figure~\ref{fig:eventrate} shows that a few hundred events are
expected in the hours before core-collapse. For the 15\,$\msun$
model, the neutrino signal exceeds $\simeq$\,100 events at $t$=$-$4~hr
and has a characteristic peak at $t\simeq -2.5$ hours, which marks the
beginning of core silicon burning. For the 30\,$\msun$ model, the
neutrino signal exceeds $\simeq$\,100 events at $t$=$-$2~hr. The number
of events then increases steadily and rapidly, leading to a cumulative
number of events that is larger than in the 15\,$\msun$ model.


For the detector background, we follow the event rates estimated in \cite{an_2016_aa} (see also  \cite{yoshida_2016_aa}) for JUNO:
$r^{on}_{Bkg}\simeq 2.66/{\rm hr}$ and $r^{off}_{Bkg}\simeq 0.16/{\rm hr}$
 in the reactor-on and
reactor-off cases respectively. In addition to reactor neutrinos, other backgrounds are due, in comparable amounts (about 1 event per day each), to geoneutrinos, cosmogenic $^8$He/$^9$Li, and accidental coincidences due to various radioactivity sources, like the natural decay chains, etc. 
For the latter, it is assumed that an effective muon veto will be in place,
see \cite{an_2016_aa} for details\footnote{Although we use detector-specific background rates, we emphasize that our results are given as a function of the forward-backward asymmetry of the data set at hand, and therefore are broadly applicable to different detector setups. See Sec. \ref{sec:method}.}. 
Roughly, a signal is detectable if the number of events
expected is at least comparable with the number of background events in the same time
interval ($N \gtrsim N_{\rm bkg}$).  Using the reactor-on background
rate, the most conservative \ps\ event rate in Figure~\ref{fig:eventrate},
and the fact that the number of signal events scales like $D^{-2}$, we
estimate that a \ps\ can be detected to a distance $D_{\rm max}
\simeq$ 1~kpc.

What nearby stars could possibly undergo core collapse in the next few decades? To answer this question, we compiled a new list of 31 core collapse \sn\ candidates; see Appendix~\ref{appendixA} and  Table~\ref{tab:presncandidates}. 
Figure \ref{fig:nearmassive2} gives an illustration of their names, positions,
distances, masses, and colors.  
Figure \ref{fig:mollweidepresn} shows the equatorial coordinate system
positions of the same stars, colored by distance bins, in a
Mollweide projection. These candidates
lie near the Galatic Plane, with clustering in directions
associated with the Orion A molecular cloud \citep{grossschedl_2019_aa} 
and the OB associations Cygnus OB2 and Carina OB1 \citep{lim_2019_aa}.
We find that for the stars in Table~\ref{tab:presncandidates} the minimum separation (i.e., the separation of a star from its nearest neighbor in the same list) is, on average, $\langle \Delta \theta \rangle\simeq
10.4^{\circ}$, and that 70\% of the candidate stars have
$\Delta \theta \lesssim 12.8^{\circ}$ (see Table \ref{tab:angularsep}).  Therefore, a sensitivity of
$\simeq$\,10$^{\circ}$ is desirable for complete disambiguation of the
progenitor with a neutrino detector.

\section{Angular Resolution and Sensitivity}
\label{sec:method}

Here we discuss the angular sensitivity of a liquid scintillator
detector for realistic numbers of \ps\ \n\ events. We consider
two cases: a well tested liquid scintillator technology
(henceforth \ls) based on Linear AlkylBenzene (LAB), as is used in SNO+
\citep{Andringa:2015tza} and envisioned for JUNO; and a
hypothetical setup where a Lithium compound is dissolved in the
scintillator for enhanced angular sensitivity (henceforth \lli), as
discussed for geoneutrino detection \citep{tanaka_2014_aa}. 
As a notation definition, let us 
assume that the
total number of events in the detector is $N = N_S + N_{\rm Bkg}$,
where $N_S$ is the number of signal events and $N_{\rm Bkg}$ is the
number of background events.

\begin{figure}
  \includegraphics[width=0.46\textwidth,angle=0]{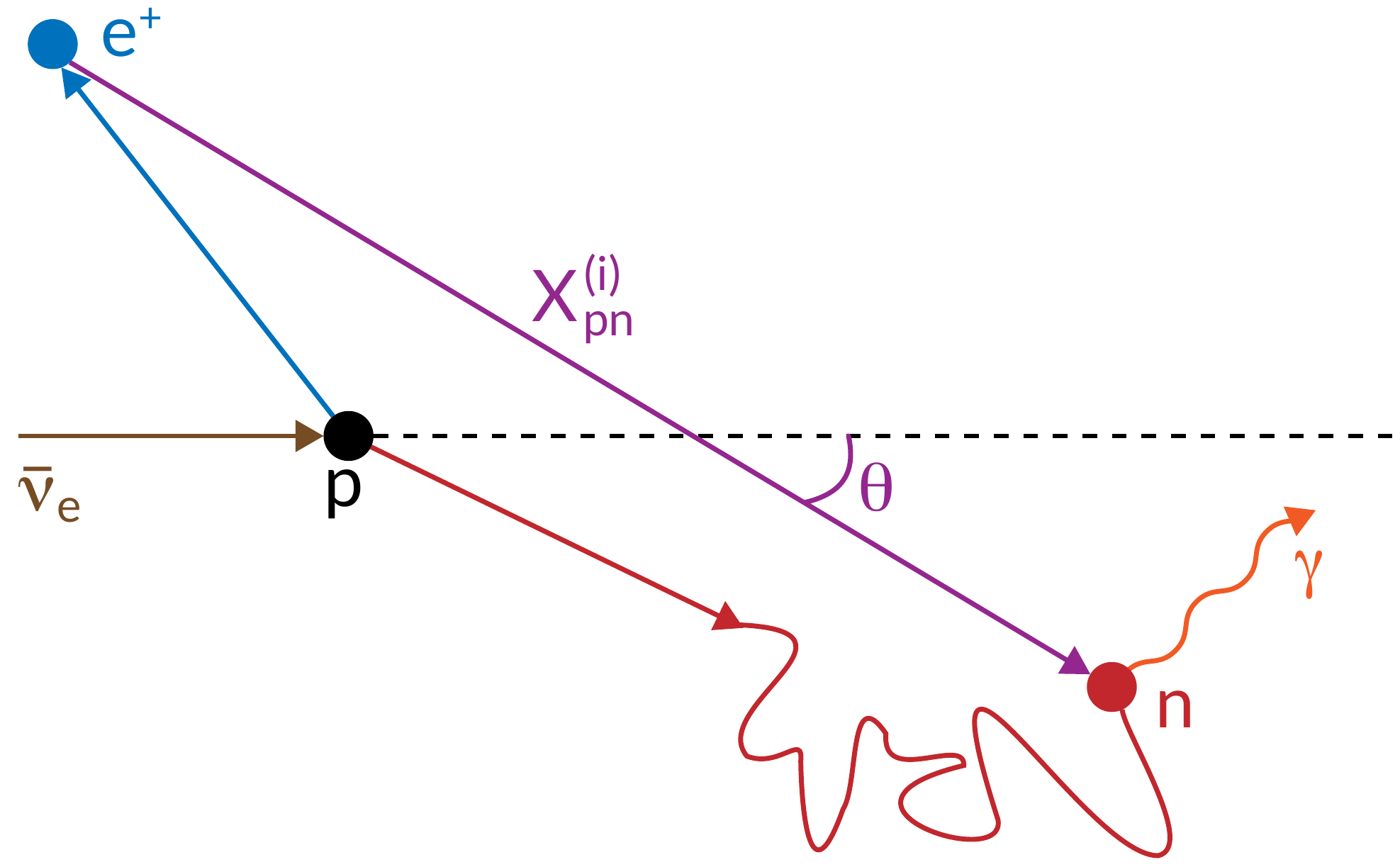}
  \caption{The geometry of Inverse Beta Decay in  liquid scintillator. 
           Shown are the incoming anti-neutrino (brown), proton (black), outgoing positron
           and its annihilation point (blue), outgoing neutron, its subsequent scattering events
           and its capture point (red), and the outgoing photon (orange).
           The vector $X^{(i)}_{pn}$ originates at the positron annihilation location
           and points in the direction of the neutron capture point. $\theta$ is the angle between $X^{(i)}_{pn}$ and the incoming neutrino momentum.
  }
\label{fig:ibd}
\end{figure}

\begin{figure}[!htb]
    \includegraphics[width=0.44\textwidth,angle=0]{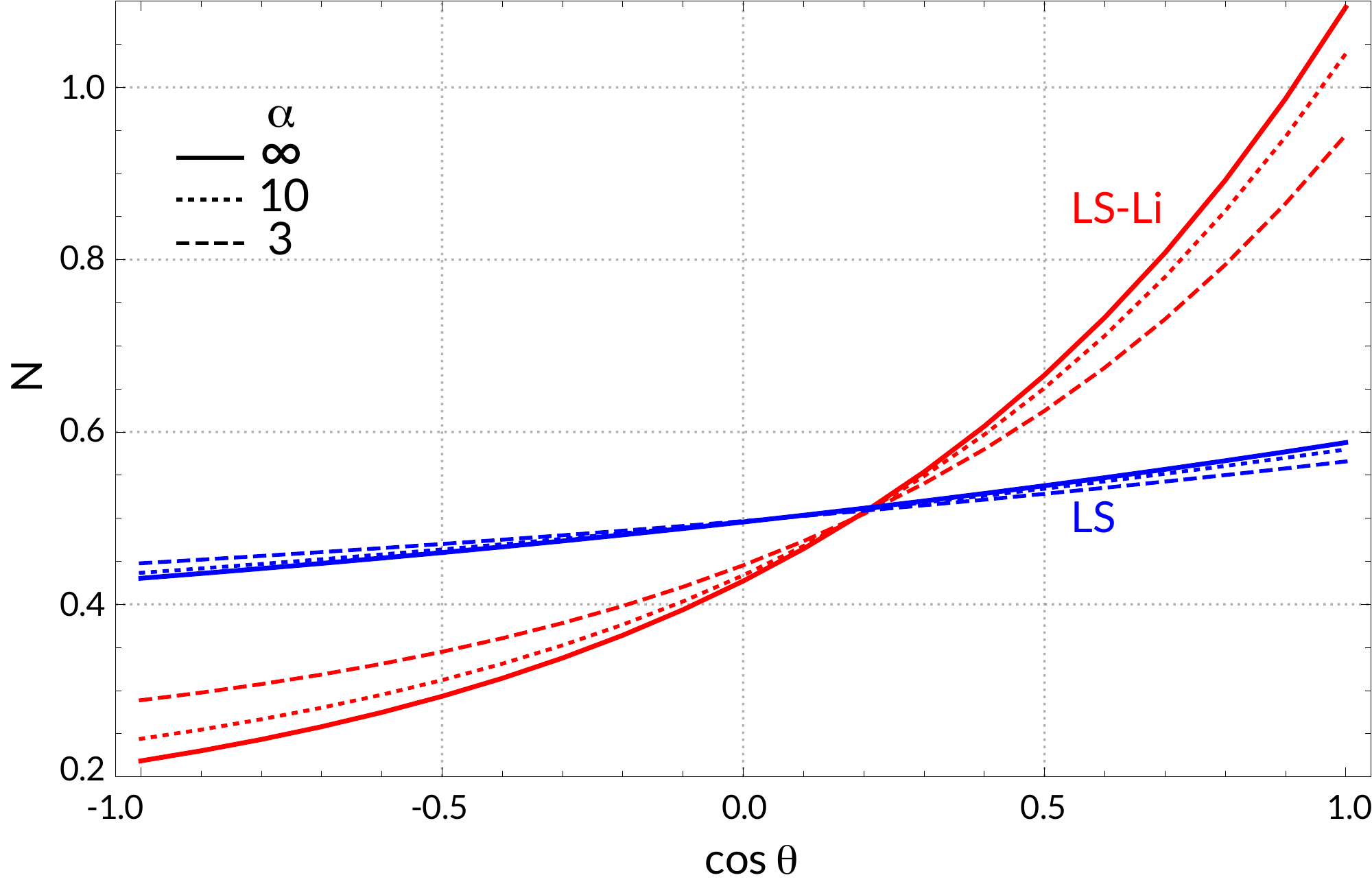}
    \caption{Normalized distributions of $\cos{\theta}$ for \ls\
             and \lli, for different values of the signal-to-background ratio,  
             $\alpha=N_S/N_{\rm Bkg}$ (numbers in legend). 
             Here, $\alpha=\infty$ means absence of background, $N_{\rm Bkg}=0$.
            }
\label{fig:cosdist}
\end{figure}

The IBD process in \ls\ is illustrated in Figure \ref{fig:ibd}. Overall, the sensitivity of this process to the direction of the incoming \n\ is moderate, with the emitted positron (neutron) momentum being slightly backward (forward)-distributed, see    
\cite{beacom_1999_aa} and \cite{vogel_1999_aa} for
a detailed overview.
%
%
%
Here, we follow the pointing method proposed and tested by
the CHOOZ collaboration \citep{apollonio_2000_aa}, which we describe briefly below. 

Let us first consider a background-free
signal, $N_{\rm Bkg}=0$. For each detected 
neutrino  $\nu_i$ ($i=1,2,$\ldots$,N$), we consider the unit vector
$\hat{X}^{(i)}_{pn}$ that originates at the positron annihilation
location and is directed towards the neutron capture point. Let $\theta$ be the angle that $\hat{X}^{(i)}_{pn}$ forms with the \n\ direction (see Figure \ref{fig:ibd}).
The unit vectors $\hat{X}^{(i)}_{pn}$
carry directional information -- albeit with some degradation due to the
neutron having to thermalize by scattering events before it can be
captured -- and possess a slightly forward distribution. The angular distributions expected for \ls\ and \lli\ are given by \cite{tanaka_2014_aa} (in the context of geoneutrinos)  in graphical form; we find that they are well reproduced by the following functions:
%
%
\begin{equation}
\begin{aligned}
& f_{LS}(\cos{\theta})    \simeq 0.2718 + 0.2238\ \exp \left( 0.345 \cos{\theta}\right) \\
& f_{LS-Li}(\cos{\theta}) \simeq 0.1230 + 0.3041\ \exp \left( 1.16 \cos{\theta} \right).
\label{twdist}
\end{aligned}
\end{equation}

Using these, one can find the forward-backward asymmetry, which is a measurable parameter:
\begin{equation}
\frac{a_0}{2}= \frac{N_F-N_B}{N_F+N_B}~.
\label{asymmetry}
\end{equation}
Here $N_F$ and $N_B$ are the numbers of events in the forward
($\theta\leq \pi/2$) and backward ($\theta >\pi/2$) direction 
respectively. 
We obtain $a_0\simeq 0.16$ for \ls, which is consistent with the distributions shown in \cite{apollonio_2000_aa}, and $a_0\simeq 0.78$ for \lli. 

Let us now generalize to the case with a non-zero background, and define  the signal-to-background ratio, $\alpha=N_S/N_{\rm Bkg}$.  For simplicity, the background is modeled as isotropic and constant in time.
Suppose that $N_S$, $\alpha$, and $a_0$ are known.
In this case, the total angular distribution of the $N$ events will be a
linear combination of two components, one for the directional signal
\begin{equation}
N_{B,S}  = \frac{N_S}{2} \left(1 - \frac{a_0}{2}\right) \qquad 
N_{F,S}  = \frac{N_S}{2} \left(1 + \frac{a_0}{2}\right),
\label{directional}
\end{equation}
and the other for the isotropic background 
\begin{equation}
N_{B,{\rm Bkg}}  = \frac{N_{\rm Bkg}}{2} \qquad
N_{F,{\rm Bkg}}  = \frac{N_{\rm Bkg}}{2}.
\label{isotopic}
\end{equation}
The two distributions have a relative weight of $\alpha$,
which yields the forward-backward asymmetry  as 
\begin{equation}
\frac{a}{2} = \frac{(N_{F,S} + N_{F,{\rm Bkg}}) - (N_{B,S} + N_{B,{\rm Bkg}})}{(N_{F,S} + N_{F,{\rm Bkg}}) + (N_{B,S} + N_{B,{\rm Bkg}})} .
\label{aeff}
\end{equation}
In the small background limit, 
$N_{\rm Bkg} \rightarrow 0$, then $\alpha \rightarrow \infty$ and $a \rightarrow a_0$.
In the large background limit 
$N_{\rm Bkg} \rightarrow \infty$, then $\alpha \rightarrow 0$ and $a \rightarrow 0$.

Figure \ref{fig:cosdist} shows the angular distribution for different
signal-to-noise ratios $\alpha$ (see Table \ref{tab:atable} for the corresponding values of $a$).  For \ls\ the $\alpha = \infty$ curve
(blue solid) is taken from Equation~(\ref{twdist}), and for \lli\ the
$\alpha = \infty$ curve (red solid) is taken from
Equation~(\ref{twdist}).  For \lli, an enhancement in the directionality
is achieved as a result of an improved reconstruction of the positron
annihilation point and a shortening of the neutron capture
range. Enhancement in the directionality decreases for \ls\ and
\lli\ as the background becomes larger.

From here on, for all cases we adopt 
an approximate linear distribution for the $N$ events in the detector:
\be
f(\cos{\theta}) = \frac{1}{2}\Big(1+a\cos{\theta}\Big)~.
\label{distribution}
\ee
This form is accurate -- yielding results that are
commensurate with those obtained from the distributions in Figure \ref{fig:cosdist} -- and it allows to describe our results as functions of the varying parameter $a$ in a general and transparent manner.  

\begin{deluxetable}{ccc}[!htb]
\tablecolumns{3}
\tablecaption{Values of $a$ for the curves in Figure \ref{fig:cosdist}.}
\label{tab:atable}
\tablehead{
\colhead{$\alpha$} &  \colhead{LS} & \colhead{LS-Li}
}
\startdata
$\infty$  & 0.1580  & 0.7820  \\
10.0      & 0.1418  & 0.7165  \\
3.0       & 0.1170  & 0.5911 \\
\enddata
\end{deluxetable}


Rigorously, $a$  depends on the \n\ energy.  We investigated the
uncertainty associated with treating $a$ as a 
(energy-independent) constant, and found it to be negligible in the
present context where larger errors are present from, for example,
uncertainties associated with modeling of the \ps\ \n\ event rates. 
In addition, the values of $a$ used in the literature for
\sn\ \ns, reactor \ns\ and geoneutrinos
\citep[e.g.,][]{apollonio_2000_aa,tanaka_2014_aa,fischer_2015_aa} vary
only by $\simeq$\,10-20\% over a wide range of energy. The values of
$a$ in Table \ref{tab:atable} for the background-free $\alpha = \infty$ 
cases are used in \cite{tanaka_2014_aa} and
\cite{fischer_2015_aa} for geoneutrinos, which have an energy range
($E \simeq$\,2-5~MeV) and spectrum that is similar to those of \ps\ \ns.

\subsection{Pointing to the progenitor location}
\label{sub:point}

For a signal of $N$ IBD events in the detector from a point source on the sky,
and therefore a set of unit vectors $\hat{X}^{(i)}_{pn}$ 
($i=1,2,\ldots,N$), an estimate of the direction to the source is given
by the average vector $\Vec{p}$ \citep{apollonio_2000_aa,fischer_2015_aa}:
\be
\Vec{p} = \frac{1}{N}\sum_{i=1}^N \hat{X}^{(i)}_{pn}~.
\label{avg}
\ee
This vector offers an immediate way to estimate the direction to the
progenitor star in the sky.  The calculation of the uncertainty in the
direction is more involved \citep{apollonio_2000_aa}, and requires
examining the statistical distribution of $\Vec{p}$, as follows.

Consider a Cartesian frame of reference where the \n\ source is on the
negative side of the $z$-axis. In the limit of very high statistics
($N \rightarrow \infty$), the averages of the $x$- and $y$- components
of the vectors $\hat{X}^{(i)}_{pn}$ vanish. The average of the $z$-
component can be found from Equation (\ref{distribution}), and is
$\langle z \rangle = a/3$.  Thus, the mean of $\Vec{p}$ is:
\be
\label{mean}
\Vec{p}_m = (0,0,|\Vec{p}|)=(0,0,a/3)~. 
\ee
For the linear distribution in Equation (\ref{distribution}), the standard deviation is
$\sigma = (\sqrt{3-a^2})/3\simeq  1/\sqrt{3}$ (where the approximation introduces a relative error of the form $a^2/6$, which is negligible in the present context).
%
For $N \gg 1$, the Central Limit Theorem applies, and the distribution 
of the three components of $\Vec{p}$ are Gaussians\footnote{
This statement (and therefore Equation (\ref{probabilitydist})) is only valid in the assumed frame of reference, which is centered at the detector, with the neutrino source being on the $z$-axis. In a generic frame of reference, the three components of $\Vec{p}$ are not statistically independent, and their probability distribution takes a more complicated form.
} 
centered at the components of $\Vec{p}_m$, 
and with standard deviations $\sigma_x =\sigma_y=\sigma_z=\sigma= 1/\sqrt{3N}$.  Hence, the
probability distribution of the vector $\Vec{p}$ is
\begin{multline}
P(p_x,p_y,p_z) = \frac{1}{\big(2 \pi \sigma^2\big)^{\frac{3}{2}}} \\ \exp\Bigg(\frac{-p_x^2 -p_y^2 -(p_z-|\Vec{p}|)^2}{2 \sigma^2}\Bigg)\;.
\label{probabilitydist}
\end{multline}

The angular uncertainty on the direction to the \sn\ progenitor is
given by the angular aperture, $\beta$, of the cone around the
vector $\Vec{p}_m$, containing a chosen fraction of the total
probability (e.g., $I= 0.68$ or $I = 0.90$):
\be
\int P(p_x,p_y,p_z)\ dp_xdp_ydp_z = I~,
\label{pintegral}
\ee
or, in spherical coordinates:
\be
\label{spintegral}
\int_0^\infty p^2 dp \int_{\cos{\beta}}^1 d \cos{\theta} \int_0^{2 \pi} d \phi\ P(p_x,p_y,p_z) = I~. 
\ee
The latter form reduces to:
\begin{multline}
\label{simpleint}
   \frac{1}{2}\Bigg[1+{\rm Erf}(k)-\cos{\beta}\ \exp{\Bigg(-k^2\sin^2{\beta}\Bigg)} \\ \bigg(1+{\rm Erf}(k\cos{\beta})\bigg)\Bigg]  = I\;,
\end{multline}
where $k = \sqrt{3N/2}~|\vec{p}|=a \sqrt{N/6}$, and 
the error function is ${\rm Erf} (z) = 2/\sqrt{\pi} \int_0^z \exp(-t^2)\ dt$.

\begin{figure}
    \includegraphics[width=0.44\textwidth,angle=0]{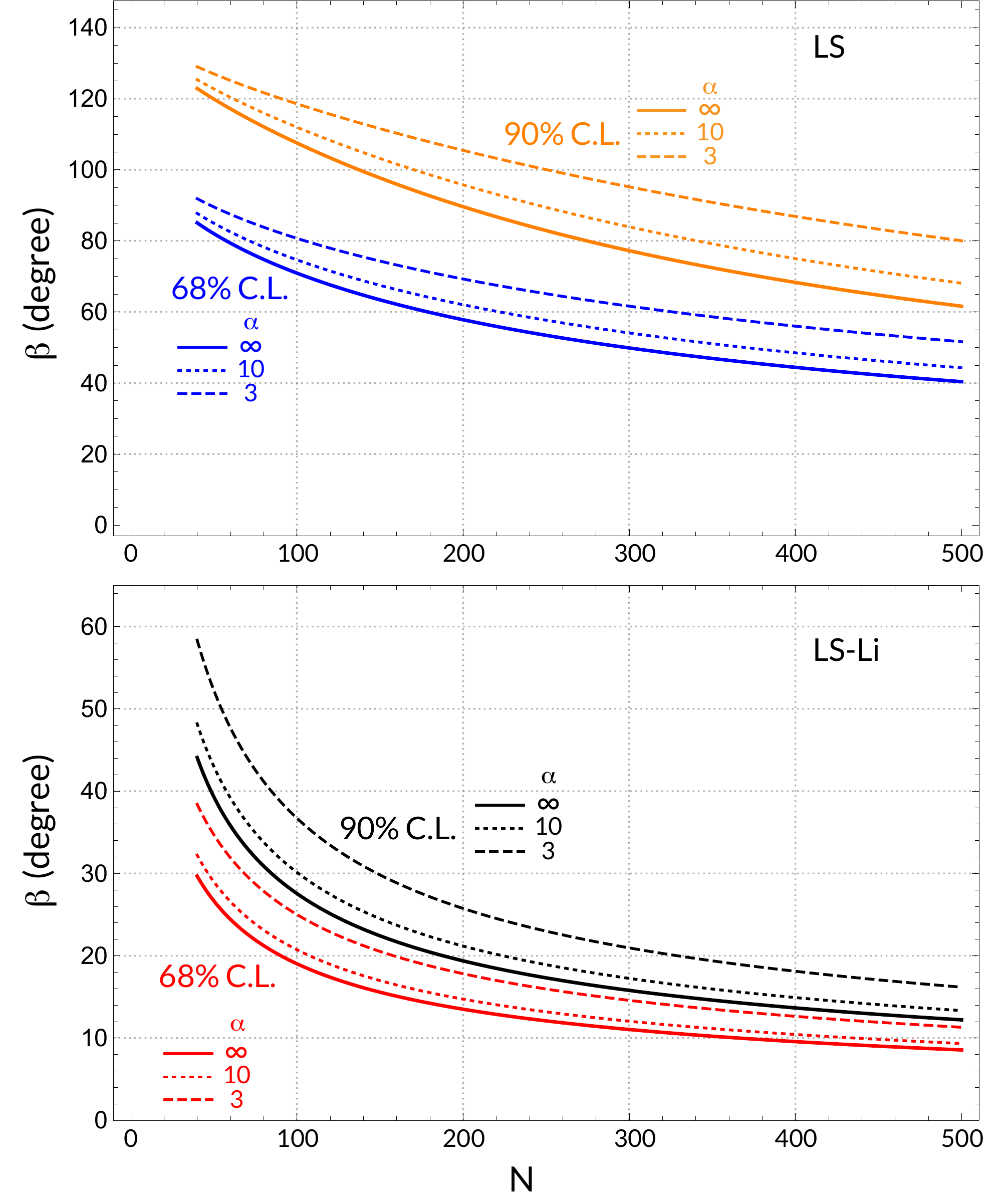}
    \caption{The angular uncertainty, $\beta$, as a function of the number of events, for \ls\ and \lli, two different confidence levels, and three values of the signal-to-background ratio, $\alpha$ (see figure legend).  
}
\label{fig:angularsens}
\end{figure}

\begin{figure}
    \includegraphics[width=0.44\textwidth,angle=0]{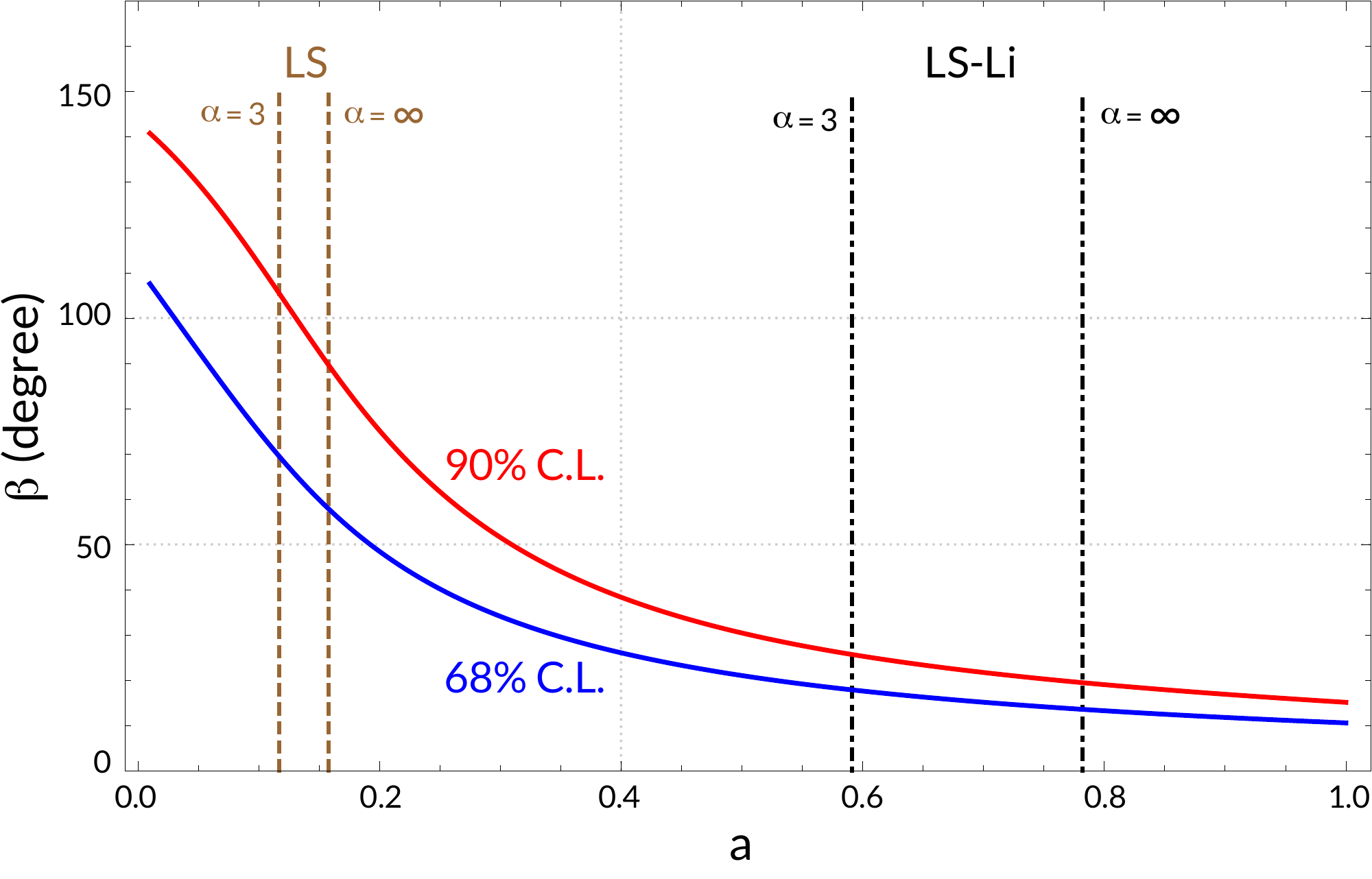}
    \caption{
    The angular uncertainty, $\beta$, as a function of the forward-backward asymmetry, $a$, for  two different confidence levels (see figure legend)  and fixed number of events, $N=200$. 
            The  vertical lines indicate the  values of $a$ corresponding to $\alpha=\infty,~3$ for \ls\ (dashed lines) and \lli\ (dot-dashed), see Table \ref{tab:atable}.  
  %
  }
\label{fig:fixedn}
\end{figure}

For a fixed value of $I$, Equation (\ref{simpleint}) can be solved
numerically to find $\beta = \beta(k, I)$, and therefore to reveal the
dependence of $\beta$ on $N$ and $a$.  Figure \ref{fig:angularsens}
shows the dependence of $\beta$ on $N$, for two confidence levels
(C.L.).  The figure illustrates the (expected) poor performance of
\ls: we have $\beta\simeq 70^\circ$ at 68\% C.L. and $N=100$,
improving to $\beta\simeq 40^\circ$ at $N=500$.  For the same C.L. and
values of $N$, \lli\ would allow an improvement in the error by nearly
a factor of 4, giving $\beta\simeq 18^\circ$ and $\beta\simeq
10^\circ$ in the two cases respectively.  The degree of improvement in
performance with increasing $a$ is shown in Figure \ref{fig:fixedn},
where $N=200$ is kept fixed.

\begin{figure*}
\centering
\subfloat [t = -4.0 hours] {\includegraphics[width=0.49\textwidth]{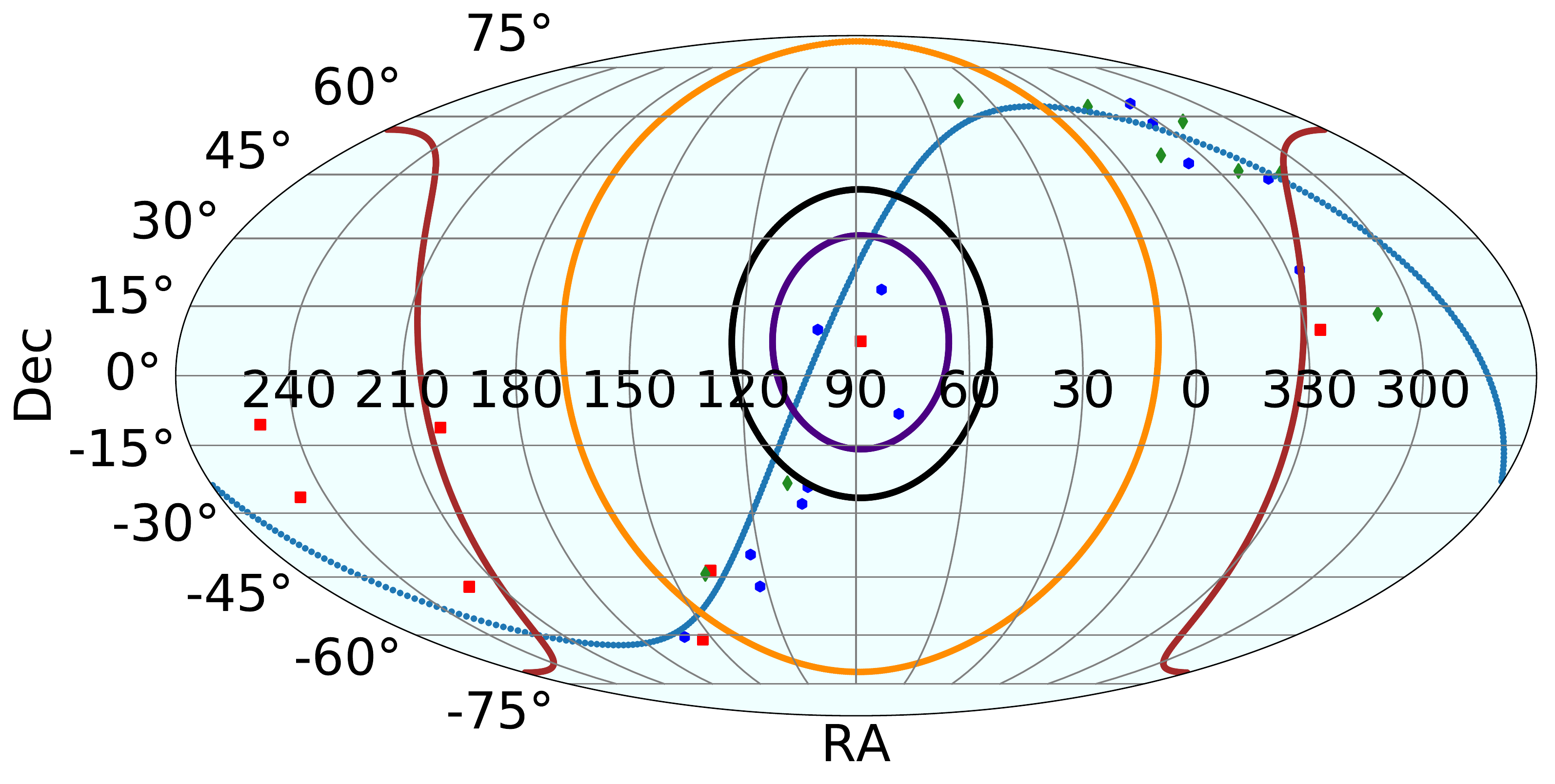}} \hfill
\subfloat [t = -4.0 hours] {\includegraphics[width=0.49\textwidth]{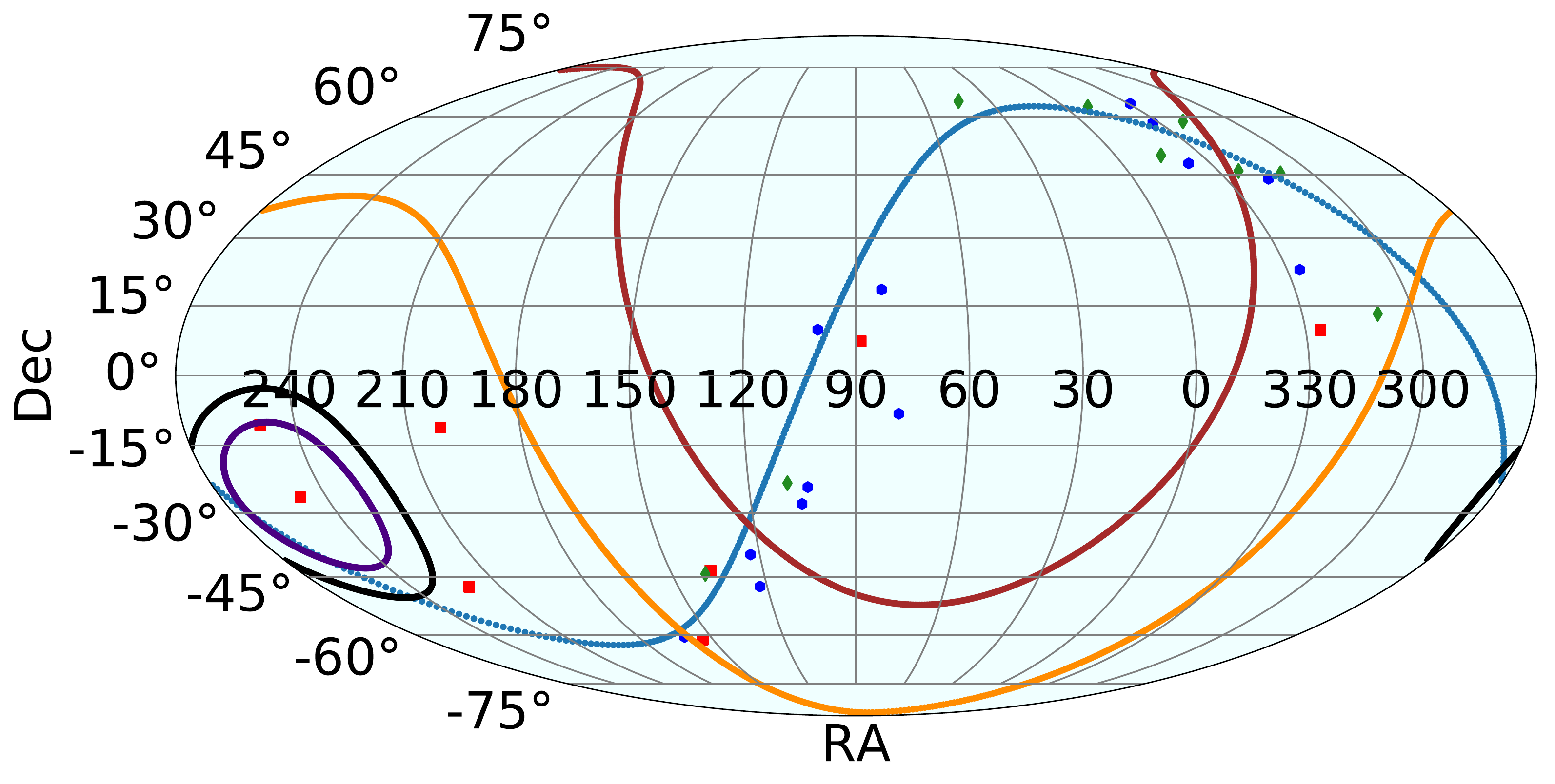}}\\
\subfloat [t = -1.0 hour] {\includegraphics[width=0.49\textwidth]{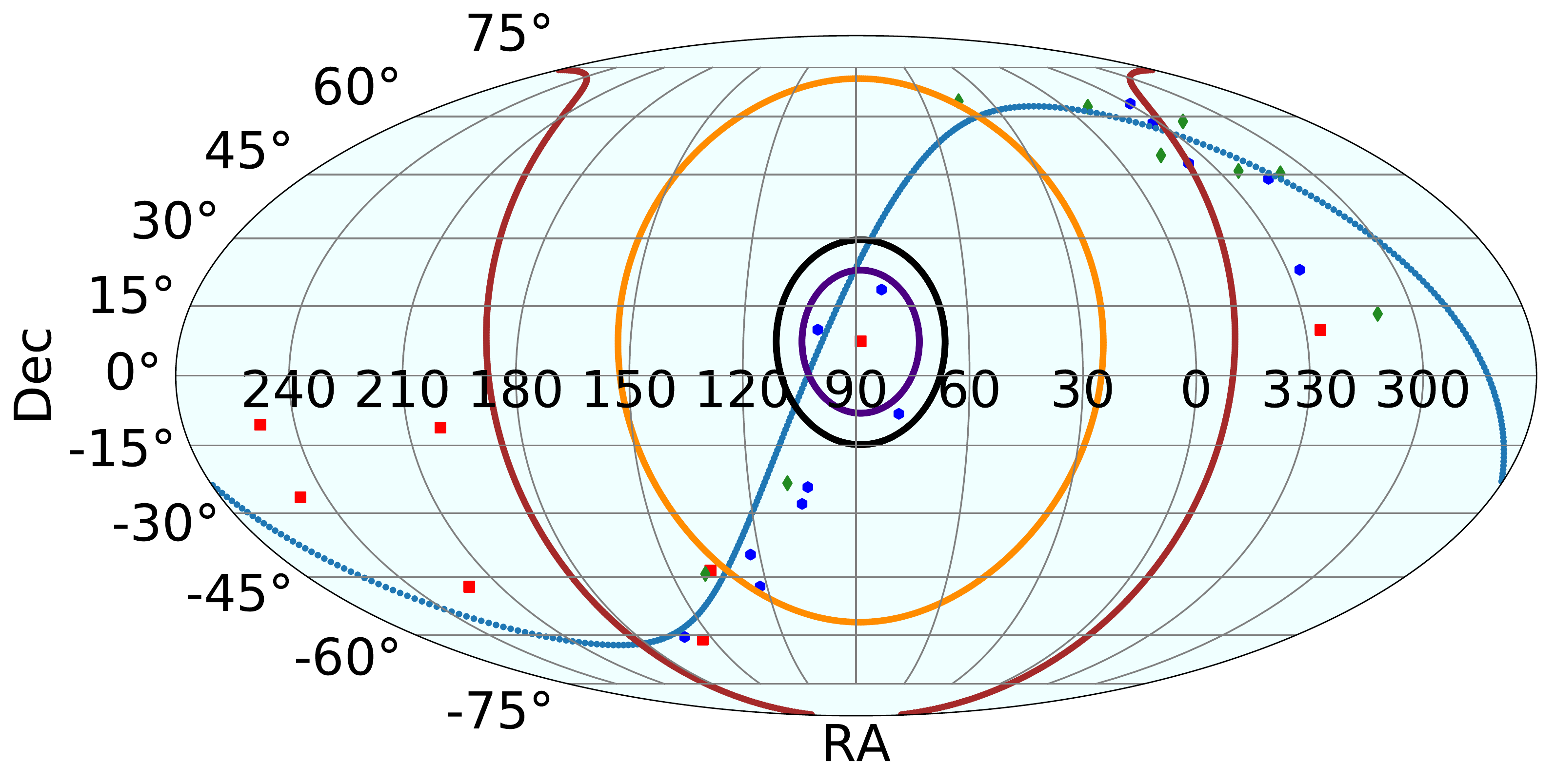}}\hfill
\subfloat [t = -1.0 hour] {\includegraphics[width=0.49\textwidth]{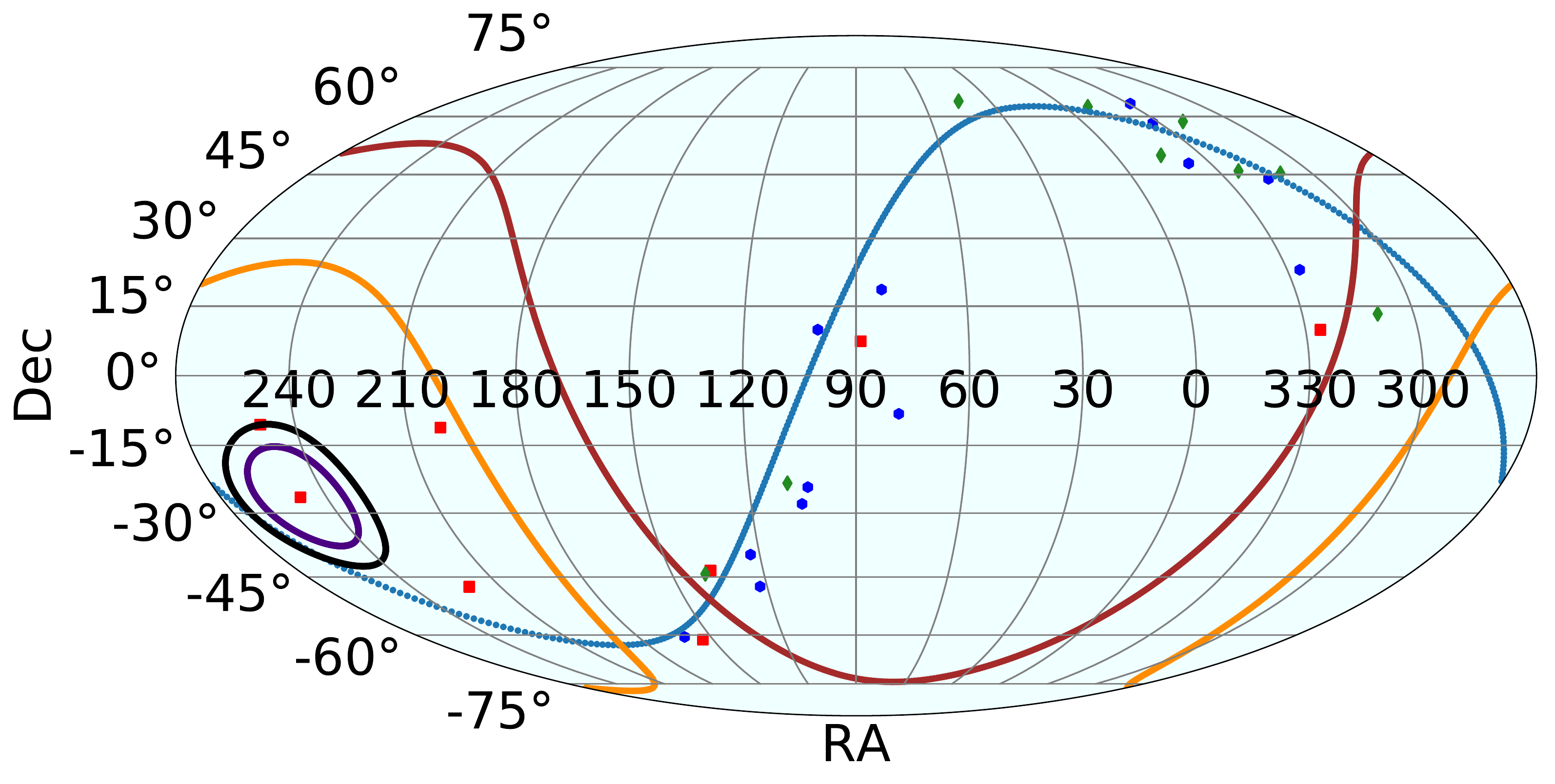}}\\
\subfloat [t = -2 minutes] {\includegraphics[width=0.49\textwidth]{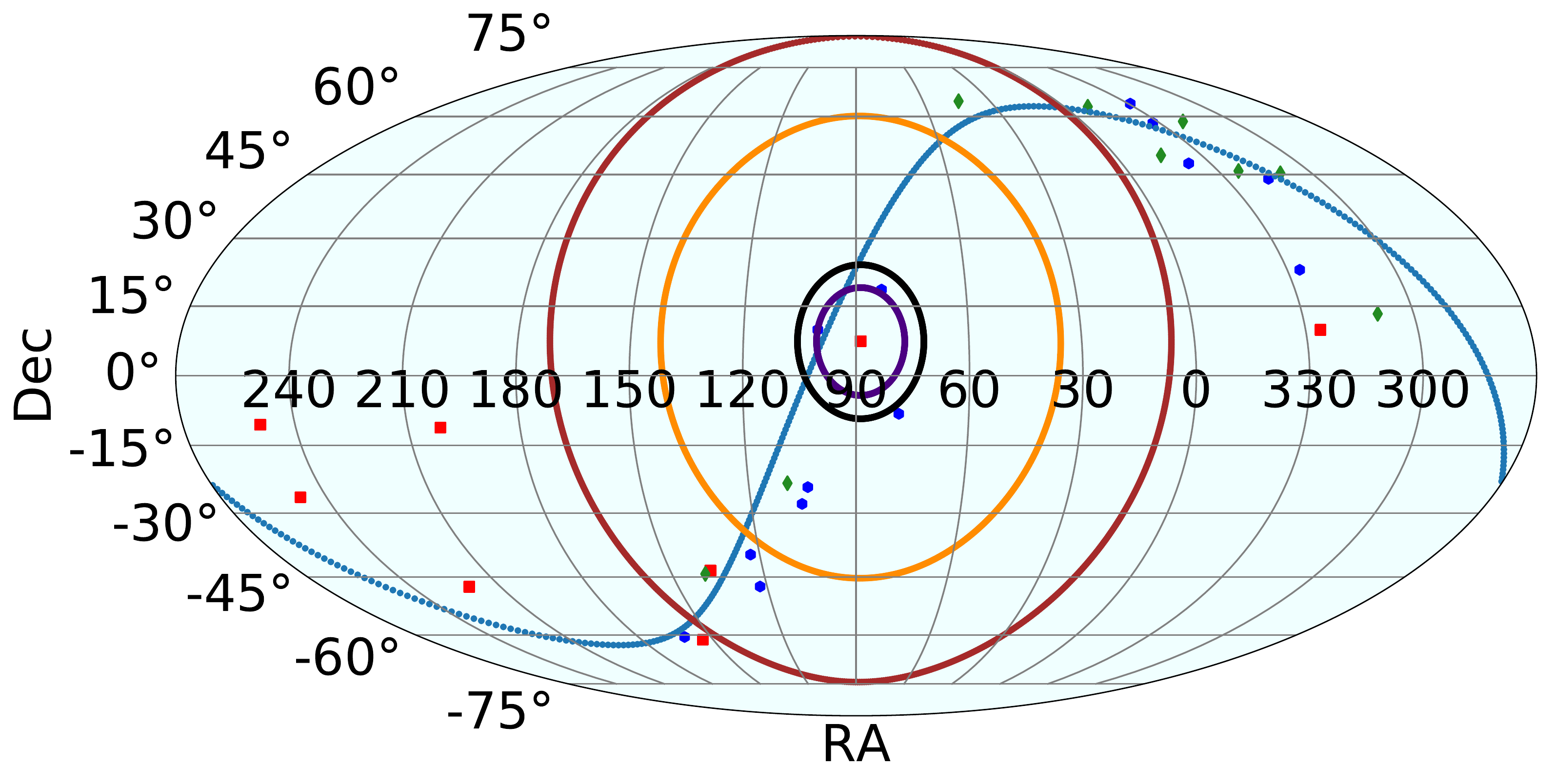}}\hfill
\subfloat [t = -2 minutes] {\includegraphics[width=0.49\textwidth]{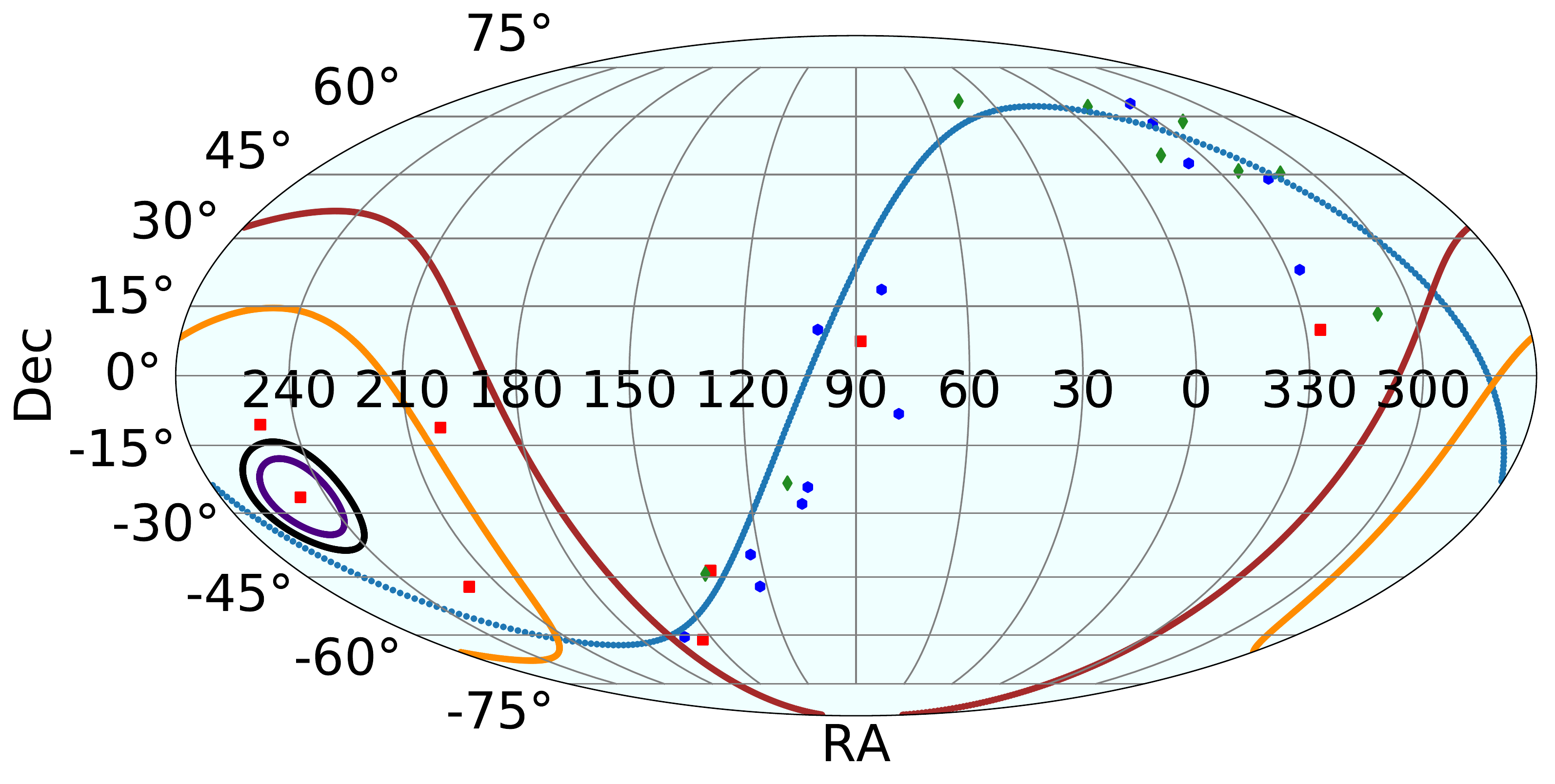}}
\vskip0.2truecm
\centering
\caption{
Angular error cones at 68\% C.L. and 90\% C.L. for \ls\ (orange and maroon contours), and \lli\ (indigo and black contours) at 4 hours, 1 hour
and 2 minutes prior to the core collapse. 
The left panels correspond to Betelgeuse ($D$=\,0.222~kpc, $M \simeq 15 \msun$);
the right panels to Antares ($D$=\,0.169~kpc, $M \simeq 15 \msun$). The
presence of background is considered in all cases according to
\cite{an_2016_aa}. The number of events is based on the model by
\cite{patton_2017_ab}.
}
\label{fig:casestudy}
\end{figure*}

\begin{figure*}
\gridline{\fig{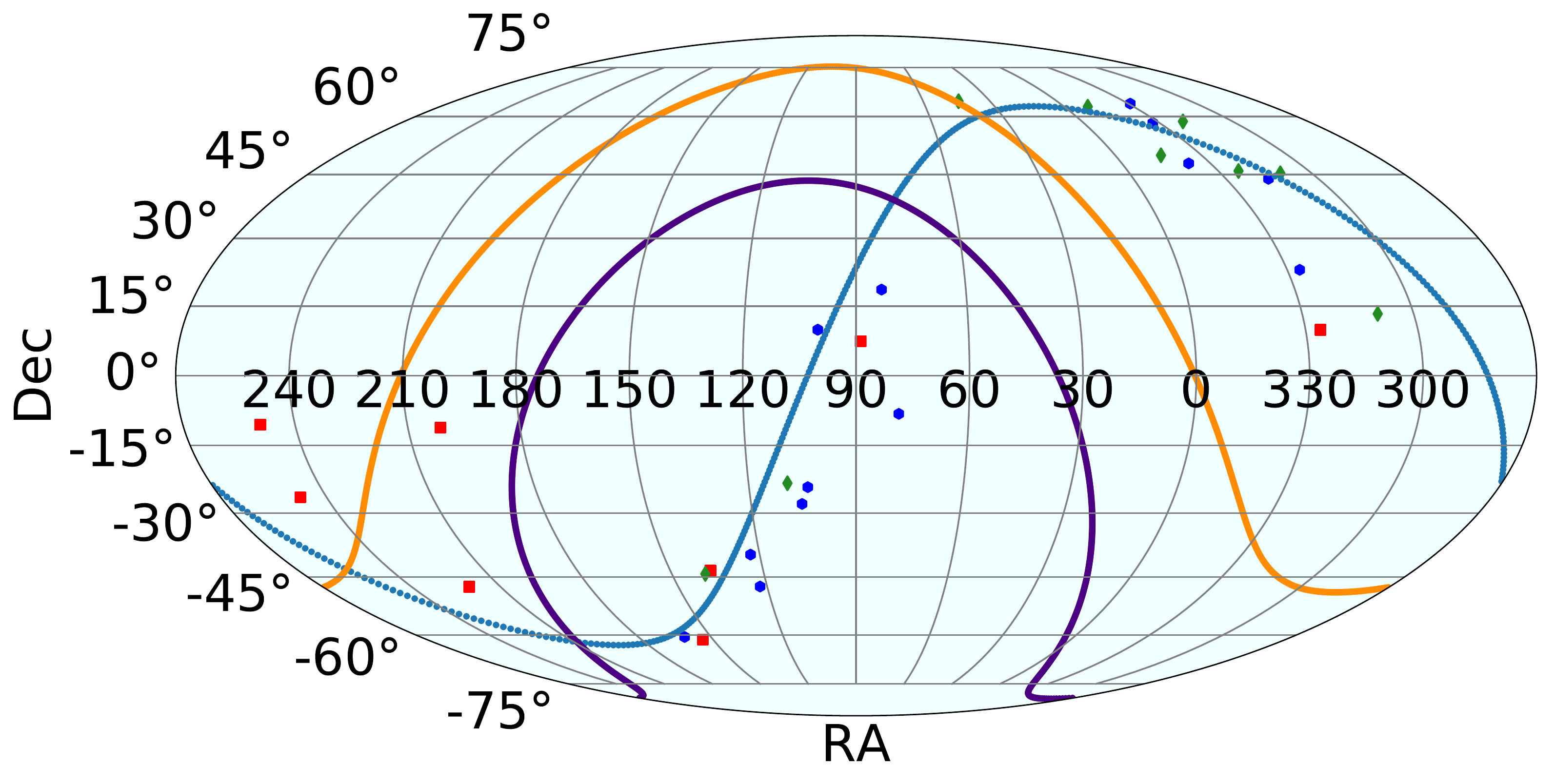}{0.49\textwidth}{t = -2.0 hours}
          \fig{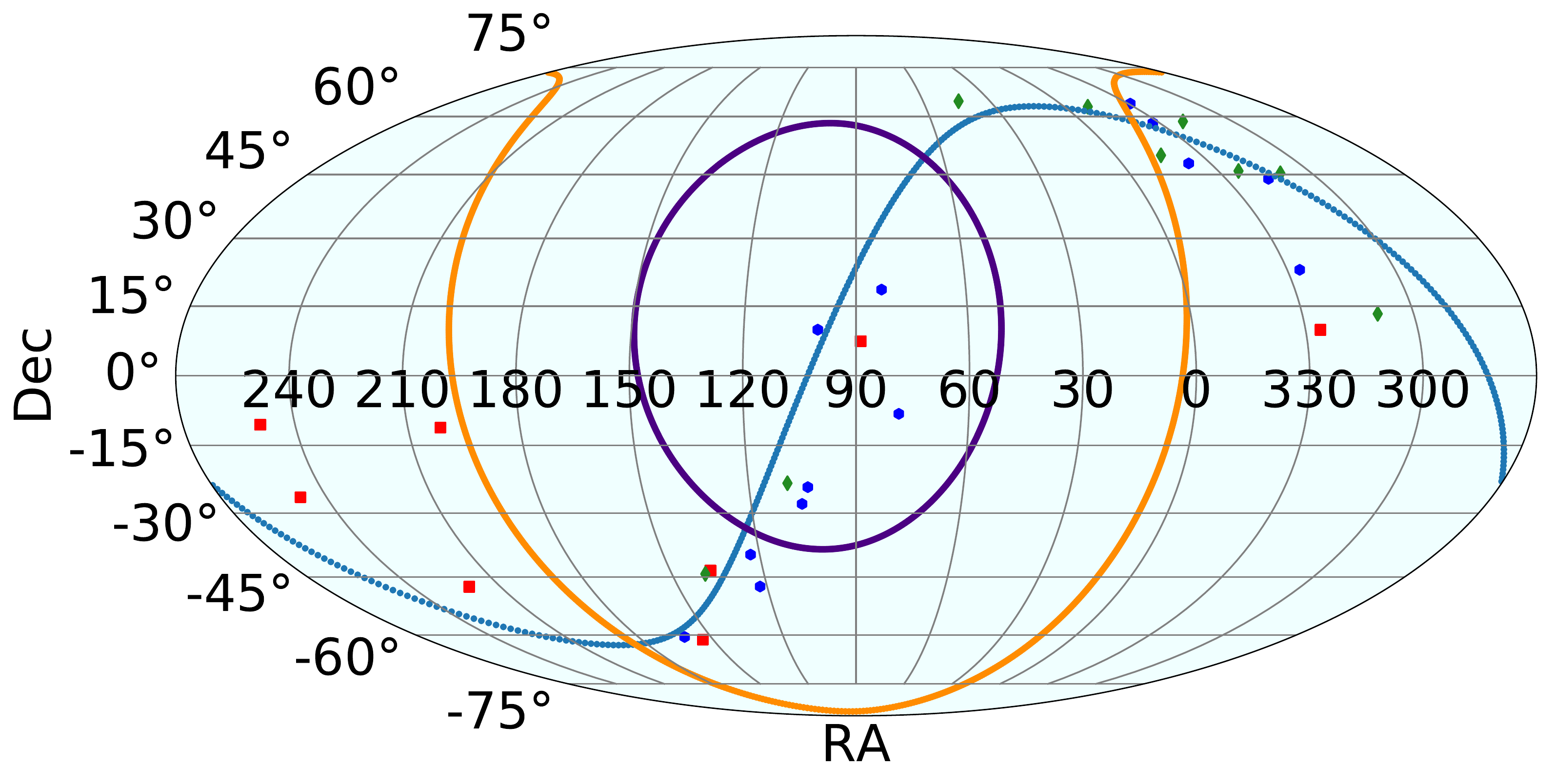}{0.49\textwidth}{t = -2.0 hours}
         }
\gridline{\fig{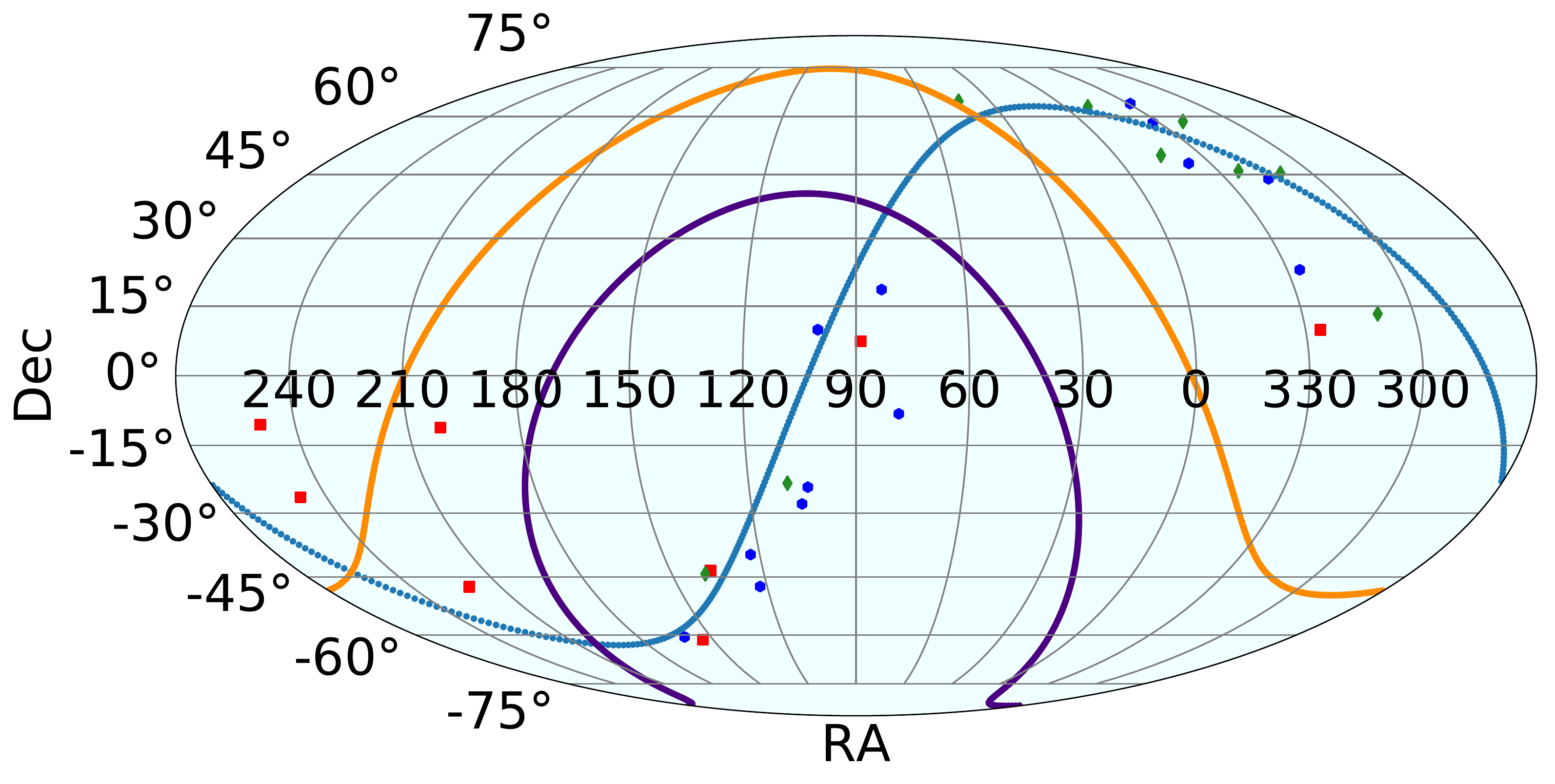}{0.49\textwidth}{t = -1.0 hour}
          \fig{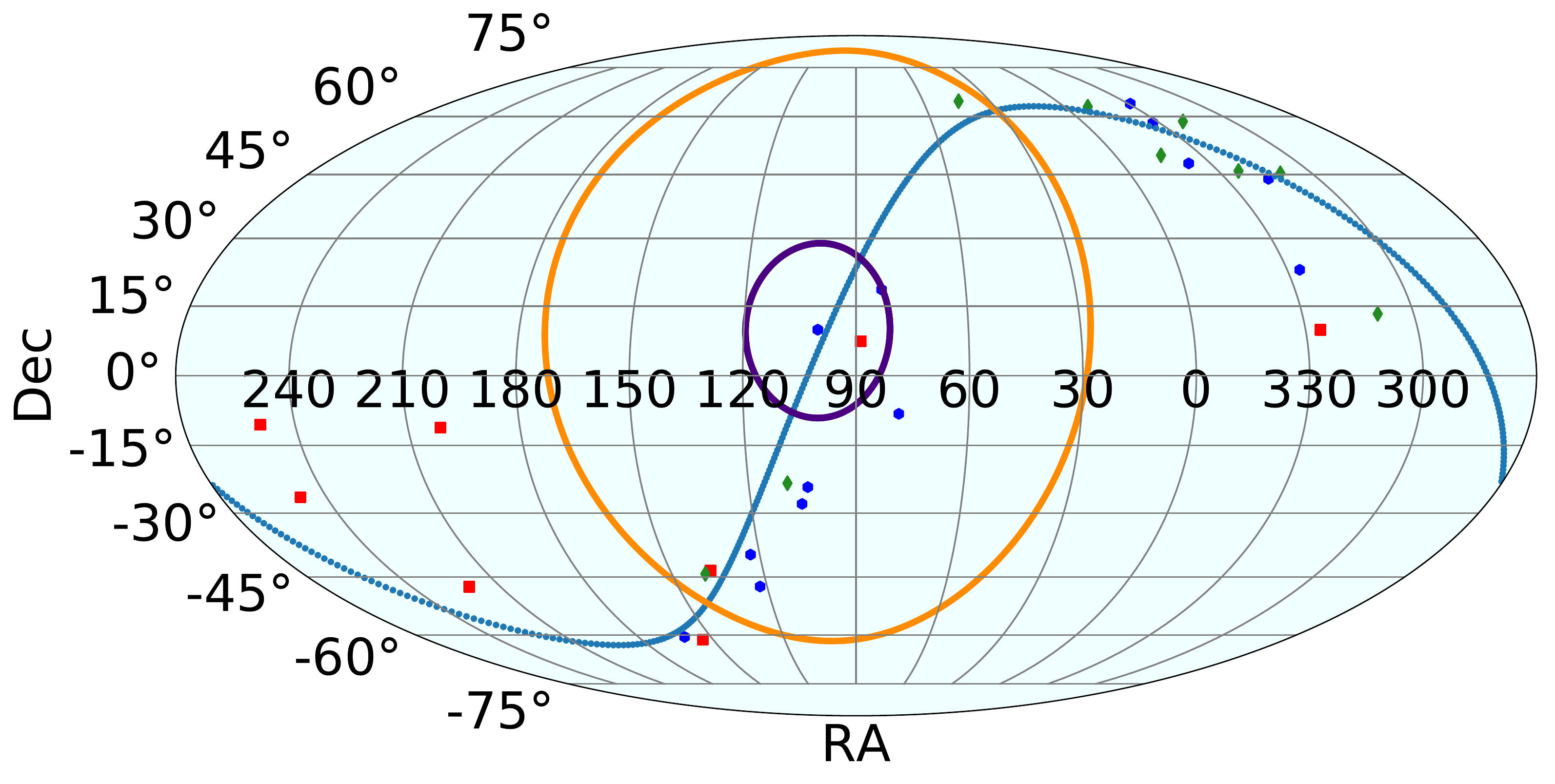}{0.49\textwidth}{t = -1.0 hour}
         }
\gridline{\fig{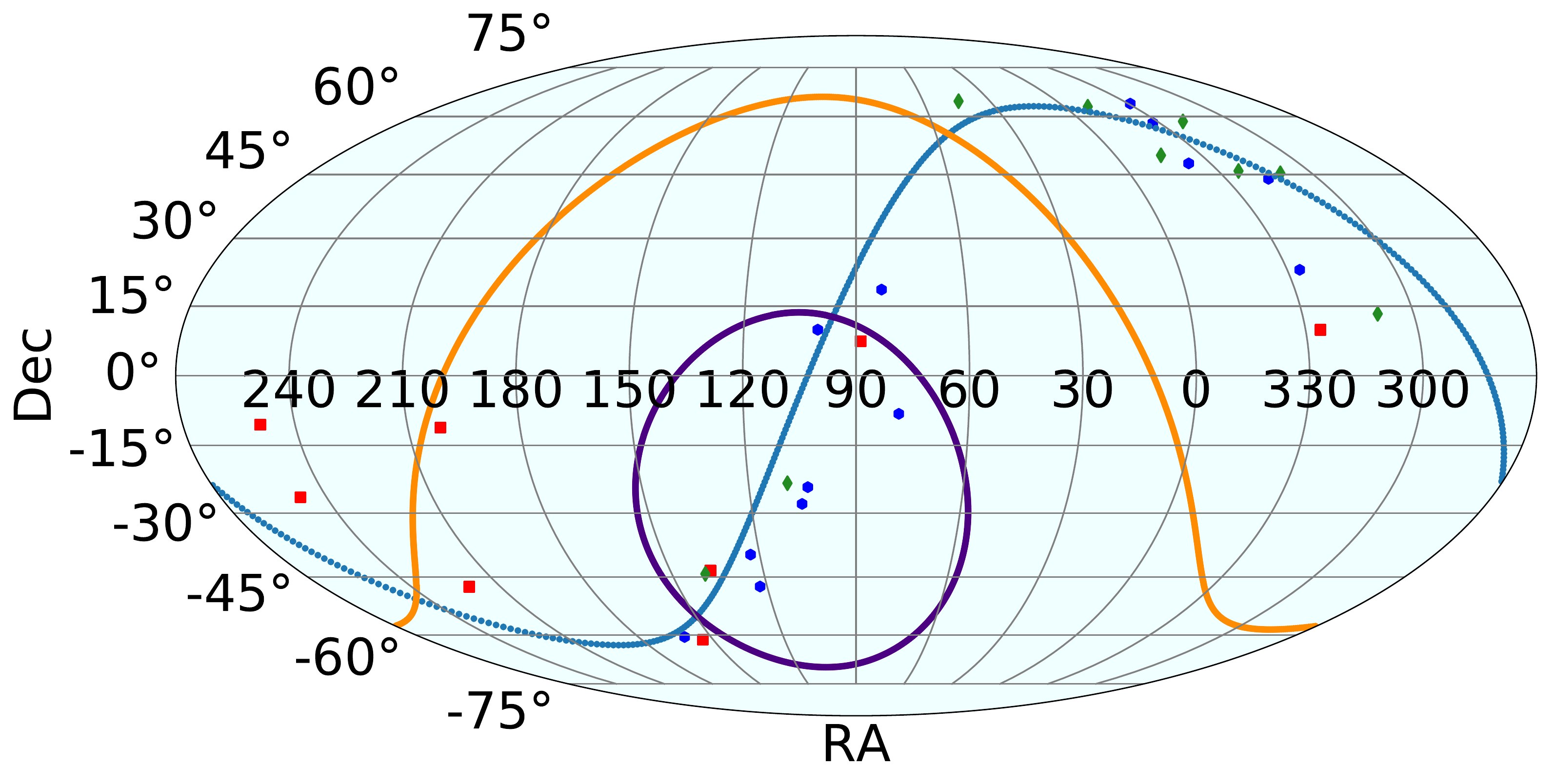}{0.49\textwidth}{t = -2 minutes}
          \fig{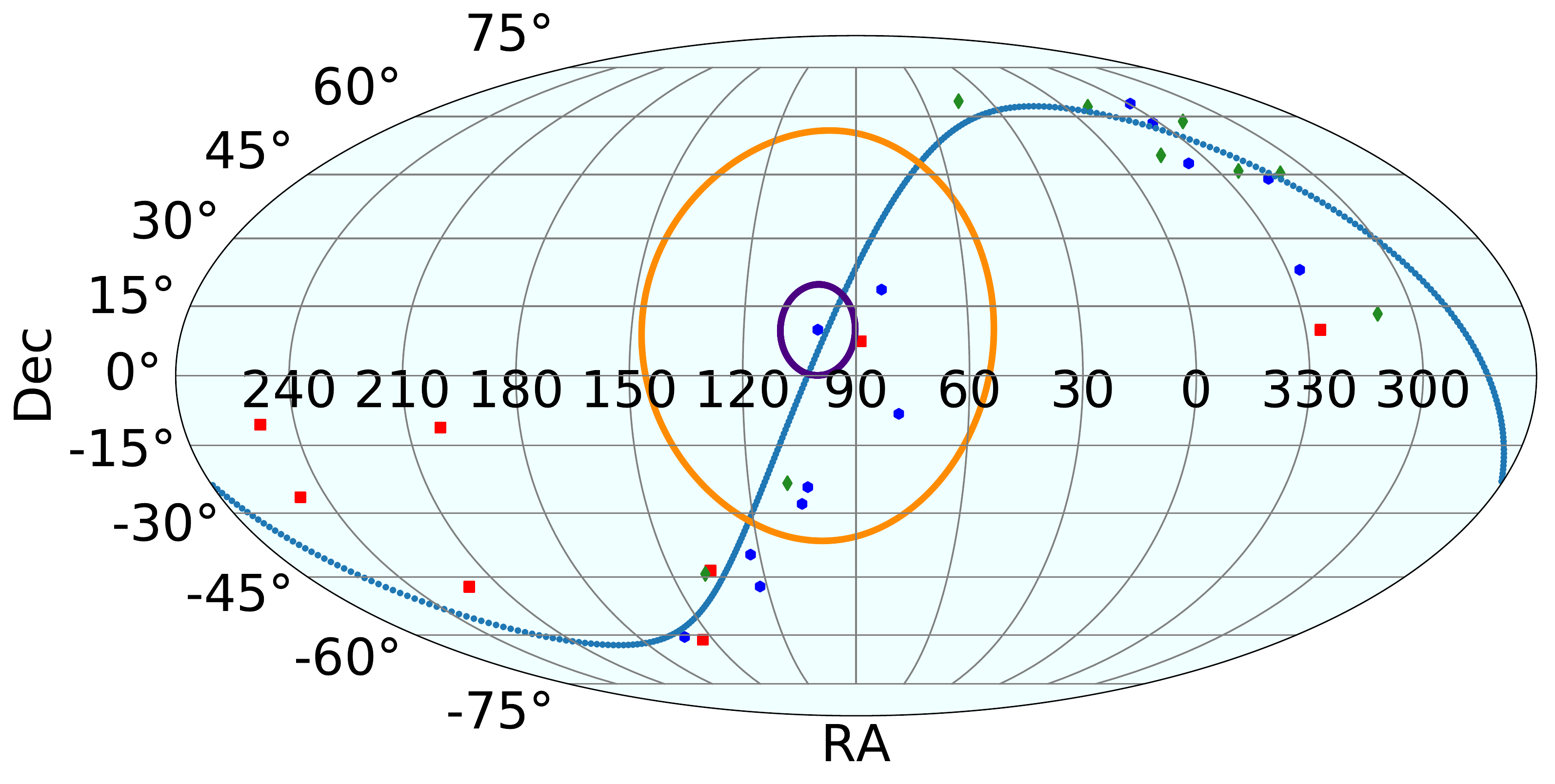}{0.49\textwidth}{t = -2 minutes}
        }
\caption{
Same as Figure~\ref{fig:casestudy}, but for 
$\sigma$ Canis Majoris (left panels, $D$=\,0.513~kpc, $M \simeq 15 \msun$)
and S Monocerotis A (right panels, $D$=\,0.282~kpc, $M \simeq 30 \msun$). Only 68\% C.L. contours are shown here, for \ls\ (orange) and \lli\ (indigo).
}
\label{fig:cs2}
\end{figure*}

In the case  of  isotropic background the mean vector, $\Vec{p}_m$,
still points in the direction of the progenitor star. That is no
longer true in the general case of anisotropic background, which would
introduce a systematic shift in the  direction of $\Vec{p}_m$. 
A naive estimate for a point-like source of background gives an (average) shift in direction by an angle $\delta \lesssim N_{Bkg}/N_S$ (valid if $N_{Bkg} \ll N_S$ and independent of $a$), corresponding to $\delta  \lesssim 4^\circ - 10^\circ$ for parameters typical of Betelgeuse (see Table \ref{tab:cs1}). A comparison with the typical values of $\beta$ indicates that the shift is probably negligible for \ls\ ($\beta \gg \delta$, typically) but might have to be considered for \lli. 
A more accurate estimate of $\delta$ 
depends on site-specific information and is beyond the scope of the
present paper.

Another source of potential uncertainty is in the site-specific  number of accidental coincidences in the detector (e.g., a coincidence between a positron from a cosmic muon decay and a neutron capture from a different process). Although here we assume a strong muon veto  \citep{an_2016_aa}, the actual performance of the veto in a realistic setting may be different and contribute to larger background levels that would negatively affect the \ps\ localization.  
See \cite{Cao:2017drk} and references therein for  technical information on realistic veto designs and their expected performance.

\section{Progenitor Identification}
\label{sec:progenid}

Attempts at progenitor identification will involve a
complex interplay of different information from different channels.
Here, we discuss a plausible, although simplified, scenario where two
essential elements are combined: (i) pointing information from a
single liquid scintillator detector, using the method in
Section \ref{sec:method}; and (ii) a rough estimate of the distance to
the star, from the comparison of the signal with
models \footnote{Circumstances that could further narrow the list of
  candidate stars include unusual electromagnetic activity from a candidate in the
  weeks or days preceding the signal, improving the distance estimate
  using data from multiple detectors, etc. }. Both these indicators
will evolve with time over the duration of the \ps\ signal, with the
list of plausible candidates becoming shorter as higher statistics are
collected in the detector. We emphasize that the goal here is not
necessarily to reduce to a single star; even reducing the list 
to a few stars (3 or 4, for example) can be useful to the 
gravitational wave and electromagnetic astronomy communities.

Consider the two case studies shown in
Figure \ref{fig:casestudy} and detailed in Tables \ref{tab:cs1} and \ref{tab:cs1a}. 
The left column refers to Betelgeuse 
and the right column to Antares, both with a time distribution
of IBD events as in Figure~\ref{fig:eventrate} for 15\,$\msun$. The
three panels show how the 68\% and 90\% C.L. angular errors decrease
with time, leading to a progressively more accurate estimate of the
position\footnote{In a realistic situation, the center of the angular
  error cone would be shifted away from the true position of the
  progenitor star by a statistical fluctuation. This effect is not
  included here.}.

For the case of \ls, at $t=-1$ hr pre-collapse, as
many as $\sim$\,10 progenitor stars are within the angular error cone,
with only a minimal improvement at later times.  Therefore, the
identification of the progenitor can not be achieved using the angular
information alone. It might be possible, however, in the presence of
a rough distance estimation from the event rate in the detector. 
In both examples, a possible upper limit of $D<0.25$ kpc (red squares
in Figure \ref{fig:casestudy}, also see Figure \ref{fig:mollweidepresn})
results in a single pre-supernova being favored.
For \lli, the angular information alone is sufficient to
favor 3-4 stars as likely progenitors already $\sim$4 hours pre-collapse. 
At $t=-1$ hr, a single progenitor can be identified in the case of Antares.

A less fortunate scenario is shown in the left panels in Figure
\ref{fig:cs2} (details in Table \ref{tab:cs2}) for $\sigma$ Canis Majoris (distance $D=0.513$ kpc). 
The number of events was calculated according to the $15\msun$ model in Figure \ref{fig:eventrate}. 
The
lower signal statistics (the number of events barely reaches 60), and
the larger relative importance of the background result in a decreased angular
sensitivity.  We find that \ls\ will only eliminate roughly half of the sky
if we use the 68\% C.L. error cone.  When combined with an approximate
distance estimate, this coarse angular information might lead to
identifying $\sim$\,10 stars as potential candidates. With \lli, the
list of candidates might be slightly shorter but a unique identification
would be very unlikely, even immediately before collapse.

A 30\,$\msun$ case is represented by the right panels in Figure
\ref{fig:cs2} (and detailed in Table \ref{tab:csa}) for S Monocerotis A (distance $D=0.282$ kpc).
An hour prior to the collapse $\simeq$\, 120 events are expected,
allowing \ls\ to shorten the progenitor list to $\simeq$ 12 stars
within the error cone at $68\%$ C.L. Whereas, \lli\ narrows the 
progenitor list down to $\simeq$\,3 stars with the same C.L. one hour prior to the collapse. 
When combined with a rough distance estimate,
the progenitor might be successfully identified.

In closing of this section, let us elaborate on the potential of estimating the distance to the star by comparing the observed \n\ event rate with models. The accuracy of such estimate depends on the uncertainty on model predictions, which in principle can be estimated from the spread in the \ps\
neutrino number luminosity from different models in the current literature that begin with the same zero-age main sequence mass.
Unfortunately, the
\ps\ models in the present literature do not allow a
reasonable direct comparison due to key, yet often undisclosed,
modelling choices made during the evolution of a stellar model (although see \cite{patton_2017_ab} for an exception). For example,
the neutrino number luminosity can change by more than an
order of magnitude due to the prescription used for mass loss
by stellar winds over the evolution of the model, the treatment
of convective boundaries, the spatial (mass) and temporal resolution
of the model over its evolution, the global conservation of energy by
the model over its evolution, the number of isotopes evolved by
the nuclear reaction network, and how nuclear burning is coupled to
the hydrodynamics (operator split versus fully coupled vs post-processing) 
especially during the advanced stages of massive star evolution.
We must conclude, therefore, that the idea to use models to place distance constraints will become realistic only in the future, after 
more progress is achieved on \ps\ emission  models.

\section{Discussion}
\label{sec:discussion}

We have demonstrated that it will be possible to use the \n\ IBD
signal at a large liquid scintillator detector to obtain an early
localization of a nearby pre-supernova ($D \lesssim 1$ kpc).  The
method we propose is robust, as it has been used successfully for
reactor neutrinos, and it is sufficiently simple that it can be
implemented during a pre-supernova signal detection.  For a detector
where the forward-backward asymmetry is about 10\% (realistic for
JUNO), and 200 events detected (also realistic at JUNO, for a star
like Betelgeuse) the angular resolution is $\beta \simeq 60^\circ$,
which is moderate, but sufficient to exclude a large number of
potential candidate progenitors.

The method has the potential to become even more sensitive if it is
used with \lli, and therefore it provides further motivation to
develop new experimental concepts in this direction. For example, 200
signal events with forward-backward asymmetry of $\sim$40\% would
result in a resolution of about $15^\circ$, and the possibility to
uniquely identify the progenitor star.

In a realistic situation, as soon as a \ps\ signal is detected with
high confidence (a few tens of candidate events), an alert with a
coarse localization information can be issued, followed by updates
with improved angular resolution in the minutes or hours leading to
the neutrino burst detection.  

Using the \cite{patton_2017_ab} \ps\ model, we find that (see Figure
\ref{fig:casestudy}) when the number of events reaches $N=100$ 
($\simeq$\,1 hour pre-collapse for Betelgeuse), the angular information is
already close to optimal, since only a minimal improvement of the
positional estimate can be gained at subsequent times.  Note, however,
that our results are conservative. According to other simulations
where the \ps\ \n\ luminosity reaches a detectable level over a time
scale of days \citep{kato_2015_aa,Guo:2019orq}, it might be possible to
detect a larger number of events, resulting in even better angular
resolutions in the last 1-2 hours before the core collapse.

It is possible that, when a nearby star reaches its final day or hours
before becoming a supernova, a new array of \n\ detectors
will be available. A large liquid scintillator experiment like the
proposed THEIA \citep{Askins:2019oqj}, which could reach 80 kt
(fiducial) mass, could observe more than $10^3$ IBD events, with an
angular resolution of at least $\sim 30^\circ$.  The resolution of
THEIA would be improved by using a
water-based liquid scintillator, where the capability to separate the
scintillation and Cherenkov light would result in enhanced pointing
ability \citep[e.g.,][]{Askins:2019oqj} for IBD, and in the
possibility to use \n-electron elastic scattering for pointing. A
subdominant, but still useful, contribution to the pointing effort --
at the level of tens of events -- will come from $\mathcal{O}(1)$ kt
liquid scintillator projects like SNO+ \citep{Andringa:2015tza} and
the Jinping Neutrino Experiment
\citep{JinpingNeutrinoExperimentgroup:2016nol}, for which the 
deep underground depth will result in very low background levels. Further activities on
directionality in scintillators are ongoing \citep[e.g.,][]{Biller:2020uoi}.
Data from elastic scattering events at water Cherenkov detectors like
SuperKamiokande \citep{simpson_2019_aa} and possibly the planned HyperKamiokande ($\mathcal
O(100)$ kt) \citep{abe_2016_aa}, 
will also contribute, despite the loss of
statistics (compared to liquid scintillator) due to the higher energy
threshold ($\sim 5 -7 $ MeV). In these detectors, a possible phase with Gadolinium dissolved in the water, like in the upcoming SuperK-Gd, \citep{Beacom:2003nk,simpson_2019_aa}, will allow better discrimination of the IBD events, resulting in an enhanced pointing potential.

In addition to new experimental scenarios, a different theoretical
panorama may be realized as well, and there might be novel avenues to
conduct fundamental science tests (e.g., searches for exotic light and
weakly interacting particles) using \ps\ \ns. 

\begin{deluxetable*}{ccccc|ccc|ccc}[!htb]
\tablecolumns{11}
\tablecaption{Parameters and results for Betelgeuse, Figure \ref{fig:casestudy}, left panels. The angular errors at a given confidence level (C.L.) are in degrees. }
\label{tab:cs1}
\tablehead{
\multicolumn{5}{c}{} & \multicolumn{3}{c}{\ls} & \multicolumn{3}{c}{\lli} \\
\colhead{Time to CC} &  \colhead{$N_{\rm Total}$} & \colhead{$N_{\rm Signal}$} & \colhead{$N_{\rm Bkg}$} & \colhead{$\alpha$} &
\colhead{$a$} & \colhead{68\% C.L.} & \colhead{90\% C.L.} & \colhead{$a$} & \colhead{68\% C.L.} & \colhead{90\% C.L.}
}
\startdata
4.0 hr & 93  & 78  & 15 & 5.20  & 0.1308 & $78.43^\circ$ & $116.17^\circ$ & 0.6610 & $23.24^\circ$ & $33.98^\circ$ \\
1.0 hr & 193 & 170 & 23 & 7.39  & 0.1374 & $63.92^\circ$ & $98.42^\circ$ & 0.6942 & $15.47^\circ$ & $22.26^\circ$ \\
2  min & 314 & 289 & 25 & 11.56 & 0.1435 & $52.72^\circ$ & $81.79^\circ$  & 0.7254 & $11.63^\circ$ & $16.67^\circ$ \\
\enddata
\end{deluxetable*}

\begin{deluxetable*}{ccccc|ccc|ccc}[!htb]
\tablecolumns{11}
\tablecaption{Parameters and results for Antares, Figure \ref{fig:casestudy}, right panels. }
\label{tab:cs1a}
\tablehead{
\multicolumn{5}{c}{} & \multicolumn{3}{c}{\ls} & \multicolumn{3}{c}{\lli} \\
\colhead{Time to CC} &  \colhead{$N_{\rm Total}$} & \colhead{$N_{\rm Signal}$} & \colhead{$N_{\rm Bkg}$} & \colhead{$\alpha$} &
\colhead{$a$} & \colhead{68\% C.L.} & \colhead{90\% C.L.} & \colhead{$a$} & \colhead{68\% C.L.} & \colhead{90\% C.L.}
}
\startdata
4.0 hr & 161 & 146 & 15 & 9.73  & 0.1414 & $66.27^\circ$ & $101.59^\circ$ & 0.7147 & $16.44^\circ$ & $23.70^\circ$ \\
1.0 hr & 333 & 310 & 23 & 13.48 & 0.1452 & $51.11^\circ$ & $79.24^\circ$  & 0.7337 & $11.16^\circ$ & $15.98^\circ$ \\
2 min  & 543 & 518 & 25 & 20.72 & 0.1488 & $41.02^\circ$ & $62.70^\circ$  & 0.7519 & $8.54^\circ$ & $12.19^\circ$ \\
\enddata
\end{deluxetable*}

\begin{deluxetable*}{ccccc|cc|cc}[!htb]
\tablecolumns{9}
\tablecaption{Parameters and results for $\sigma$ Canis Majoris, Figure \ref{fig:cs2}, left panels. }
\label{tab:cs2}
\tablehead{
\multicolumn{5}{c}{} & \multicolumn{2}{c}{\ls} & \multicolumn{2}{c}{\lli} \\
\colhead{Time to CC} &  \colhead{$N_{\rm Total}$} & \colhead{$N_{\rm Signal}$} & \colhead{$N_{\rm Bkg}$} & \colhead{$\alpha$} &
\colhead{$a$} & \colhead{68 \% C.L.} & \colhead{$a$} & \colhead{68 \% C.L.}
}
\startdata
2.0 hr & 31 & 11 & 20 & 0.55 & 0.0553 & $103.28^\circ$ & 0.2797 & $71.43^\circ$ \\
1.0 hr & 36 & 13 & 23 & 0.56 & 0.0560 & $102.54^\circ$ & 0.2829 & $68.32^\circ$ \\
2 min  & 58 & 33 & 25 & 1.32 & 0.0887 & $93.56^\circ$  & 0.4484 & $41.57^\circ$ \\
\enddata
\end{deluxetable*}

\begin{deluxetable*}{ccccc|cc|cc}[!htb]
\tablecolumns{9}
\tablecaption{Parameters and results for S Monocerotis A, Figure \ref{fig:cs2}, right panels. }
\label{tab:csa}
\tablehead{
\multicolumn{5}{c}{} & \multicolumn{2}{c}{\ls} & \multicolumn{2}{c}{\lli} \\
\colhead{Time to CC} &  \colhead{$N_{\rm Total}$} & \colhead{$N_{\rm Signal}$} & \colhead{$N_{\rm Bkg}$} & \colhead{$\alpha$} &
\colhead{$a$} & \colhead{68 \% C.L.} & \colhead{$a$} & \colhead{68 \% C.L.}
}
\startdata
2.0 hr & 44  & 24 & 20 & 1.20  & 0.0850 & $96.53^\circ$ & 0.4300 & $48.26^\circ$ \\
1.0 hr & 141 & 118 & 23 & 5.13  & 0.1305 & $71.60^\circ$ & 0.6596 & $19.00^\circ$ \\
2 min  & 420 & 395 & 25 & 15.80 & 0.1466 & $46.28^\circ$ & 0.7413 & $9.84^\circ$ \\
\enddata
\end{deluxetable*}

\acknowledgements
We are grateful to S. Borthakur and L. M. Thomas for fruitful discussions. 
We acknowledge funding from the National Science Foundation grant number PHY-1613708.
This research was also supported at ASU by the NSF under grant PHY-1430152 for the 
Physics Frontier Center “Joint Institute for Nuclear Astrophysics—Center for the Evolution of the Elements” (JINA-CEE).
This research made extensive use of the SIMBAD Astronomical Database and SAO/NASA Astrophysics Data System (ADS).

\software{
\texttt{matplotlib} \citep{hunter_2007_aa},
\texttt{NumPy} \citep{der_walt_2011_aa}, and 
Wolfram Mathematica version 12.0.
}

\clearpage

\appendix
\restartappendixnumbering

\section{Pre-Supernova Candidates}
\label{appendixA}

Table \ref{tab:presncandidates} compiles a list of 31 red and blue
core-collapse supernova progenitors within 1~kpc that have both
distance and mass estimates.  Table \ref{tab:presncandidates} gives
the star number (sorted by distance), Henry Draper (HD) catalog
number, common name, constellation, distance, mass, J2000 right
ascension (RA) and J2000 declination (Dec).  For stars with multiple
distance measurements, precedence is given to distances provided by the
\cite{gaia-collaboration_2018_aa}, \cite{van-leeuwen_2007_aa}, and
individual determinations, in this order.  Earlier compilations
\citep[e.g.,][]{nakamura_2016_aa} considered only red supergiant
progenitors and did not require a mass estimate.

Table~\ref{tab:angularsep} lists the angular distance $\Delta \theta$
of each star to its nearest neighbor.  Table~\ref{tab:angularsep}
gives the star number, HD catalog and common name, the minimum angular
separation between the star and its nearest neighbor, the HD catalog
and common name of the nearest neighbor, and the star number of the
nearest neighbor. The RA and Dec for each star is taken from Table
\ref{tab:presncandidates} when calculating angular separations.

\begin{deluxetable*}{clcccrcl}[!htb]
\tablecolumns{8}
\tablecaption{Candidate Pre-supernova Stars.}
\label{tab:presncandidates}
\tablehead{
\colhead{N} & \colhead{Catalog Name} &  \colhead{Common Name} & \colhead{Constellation} &
\colhead{Distance (kpc)} &  \colhead{Mass (M$_{\odot}$)} & \colhead{RA} & \colhead{Dec}
}
\startdata
1  & HD 116658 & Spica/$\alpha$ Virginis & Virgo & $0.077 \pm 0.004$ \tablenotemark{a} & $11.43^{+1.15}_{-1.15}$ \tablenotemark{b} & 13:25:11.58 & $-$11:09:40.8 \\
2  & HD 149757 & $\zeta$ Ophiuchi & Ophiuchus & $0.112 \pm 0.002$ \tablenotemark{a} & $20.0$ \tablenotemark{g} & 16:37:09.54 & $-$10:34:01.53 \\
3  & HD 129056 & $\alpha$ Lupi & Lupus & $0.143 \pm 0.003$ \tablenotemark{a} & $10.1^{+1.0}_{-1.0}$ \tablenotemark{f} & 14:41:55.76 & $-$47:23:17.52 \\
4  & HD 78647 & $\lambda$ Velorum & Vela & $0.167 \pm 0.003$ \tablenotemark{a} & $7.0^{+1.5}_{-1.0}$\  \tablenotemark{h} & 09:07:59.76 & $-$43:25:57.3 \\
5  & HD 148478 & Antares/$\alpha$ Scorpii & Scorpius & $0.169 \pm 0.030$ \tablenotemark{a} & $11.0-14.3$\  \tablenotemark{l} & 16:29:24.46 & $-$26:25:55.2 \\
6  & HD 206778 & $\epsilon$ Pegasi & Pegasus & $0.211 \pm 0.006$ \tablenotemark{a} & $11.7^{+0.8}_{-0.8}$\  \tablenotemark{f} & 21:44:11.16 & +09:52:30.0 \\
7  & HD 39801 & Betelgeuse/$\alpha$ Orionis & Orion & $0.222 \pm 0.040$ \tablenotemark{d} & $11.6^{+5.0}_{-3.9}$\  \tablenotemark{m} & 05:55:10.31 & +07:24:25.4 \\
8  & HD 89388 & q Car/V337 Car & Carina & $0.230 \pm 0.020$ \tablenotemark{c} & $6.9^{+0.6}_{-0.6}$ \tablenotemark{f} & 10:17:04.98 & $-$61:19:56.3 \\
9  & HD 210745 & $\zeta$ Cephei & Cepheus & $0.256 \pm 0.006$ \tablenotemark{c}& $10.1^{+0.1}_{-0.1}$ \tablenotemark{f} & 22:10:51.28 & +58:12:04.5 \\
10 & HD 34085 & Rigel/$\beta$ Orion & Orion & $0.264 \pm 0.024$ \tablenotemark{a} & $21.0^{+3.0}_{-3.0}$ \tablenotemark{j} & 05:14:32.27 & $-$08:12:05.90\\
11 & HD 200905 & $\xi$ Cygni & Cygnus & $0.278 \pm 0.029$ \tablenotemark{c} & $8.0$ \tablenotemark{r} & 21:04:55.86 & +43:55:40.3 \\
12 & HD 47839 & S Monocerotis A & Monoceros & $0.282 \pm 0.040$ \tablenotemark{a} & $29.1$ \tablenotemark{i} & 06:40:58.66 & +09:53:44.71 \\
13 & HD 47839 & S Monocerotis B & Monoceros & $0.282 \pm 0.040$ \tablenotemark{a} & $21.3$ \tablenotemark{i} & 06:40:58.57 & +09:53:42.20 \\
14 & HD 93070 & w Car/V520 Car & Carina & $0.294 \pm 0.023$ \tablenotemark{c} & $7.9^{+0.1}_{-0.1}$ \tablenotemark{f} & 10:43:32.29 & $-$60:33:59.8 \\
15 & HD 68553 & NS Puppis & Puppis & $0.321 \pm 0.032$ \tablenotemark{c} & $9.7$ \tablenotemark{f} & 08:11:21.49 & $-$39:37:06.8 \\
16 & HD 36389 & CE Tauri/119 Tauri & Taurus & $0.326 \pm 0.070$ \tablenotemark{c} & $14.37^{+2.00}_{-2.77}$\  \tablenotemark{k} & 05:32:12.75 & +18:35:39.2 \\
17 & HD 68273 & $\gamma^2$ Velorum & Vela & $0.342 \pm 0.035$ \tablenotemark{a} & $9.0^{+0.6}_{-0.6}$ \tablenotemark{o} & 08:09:31.95 & $-$47:20:11.71 \\
18 & HD 50877 & $o^1$ Canis Majoris & Canis Major & $0.394 \pm 0.052$ \tablenotemark{c} & $7.83^{+2.0}_{-2.0}$ \tablenotemark{f} & 06:54:07.95 & $-$24:11:03.2 \\
19 & HD 207089 & 12 Pegasi & Pegasus & $0.415 \pm 0.031$ \tablenotemark{c} & $6.3^{+0.7}_{-0.7}$ \tablenotemark{f} & 21:46:04.36 & +22:56:56.0 \\
20 & HD 213310 & 5 Lacertae & Lacerta & $0.505 \pm 0.046$ \tablenotemark{a} & $5.11^{+0.18}_{-0.18}$ \tablenotemark{q} & 22:29:31.82 & +47:42:24.8 \\
21 & HD 52877 & $\sigma$ Canis Majoris & Canis Major & $0.513 \pm 0.108$ \tablenotemark{c} & $12.3^{+0.1}_{-0.1}$ \tablenotemark{f} & 07:01:43.15 & $-$27:56:05.4 \\
22 & HD 208816 & VV Cephei & Cepheus & $0.599 \pm 0.083$ \tablenotemark{c} & $10.6^{+1.0}_{-1.0}$ \tablenotemark{f} & 21:56:39.14 & +63:37:32.0 \\
23 & HD 196725 & $\theta$ Delphini & Delphinus & $0.629 \pm 0.029$ \tablenotemark{c} & $5.60^{+3.0}_{-3.0}$ \tablenotemark{n} & 20:38:43.99 & +13:18:54.4 \\
24 & HD 203338 & V381 Cephei & Cepheus & $0.631 \pm 0.086$ \tablenotemark{c} & $12.0$ \tablenotemark{s} & 21:19:15.69 & +58:37:24.6 \\
25 & HD 216946 & V424 Lacertae & Lacerta & $0.634 \pm 0.075$ \tablenotemark{c} & $6.8^{+1.0}_{-1.0}$ \tablenotemark{p} & 22:56:26.00 & +49:44:00.8 \\
26 & HD 17958 & HR 861 & Cassiopeia & $0.639 \pm 0.039$ \tablenotemark{c} & $9.2^{+0.5}_{-0.5}$ \tablenotemark{f} & 02:56:24.65  & +64:19:56.8 \\ 
27 & HD 80108 & HR 3692 & Vela & $0.650 \pm 0.061$ \tablenotemark{c} & $12.1^{+0.2}_{-0.2}$ \tablenotemark{f} & 09:16:23.03 & -44:15:56.6 \\
28 & HD 56577 & 145 Canis Major & Canis Major & $0.697 \pm 0.078$ \tablenotemark{c} & $7.8^{+0.5}_{-0.5}$ \tablenotemark{f} & 07:16:36.83 & $-$23:18:56.1 \\
29 & HD 219978 & V809 Cassiopeia & Cassiopeia & $0.730 \pm 0.074$ \tablenotemark{c} & $8.3^{+0.5}_{-0.5}$ \tablenotemark{f} & 23:19:23.77 & +62:44:23.2 \\
30 & HD 205349 & HR 8248 & Cygnus & $0.746 \pm 0.039$ \tablenotemark{c} & $6.3^{+0.7}_{-0.7}$ \tablenotemark{f} & 21:33:17.88 & +45:51:14.5 \\
31 & HD 102098 & Deneb/$\alpha$ Cygni & Cygnus & $0.802 \pm 0.066$ \tablenotemark{e} & $19.0^{+4.0}_{-4.0}$ \tablenotemark{e} & 20:41:25.9 & +45:16:49.0 \\
\enddata
\tablecomments{
$^a$\cite{van-leeuwen_2007_aa},
$^b$\cite{tkachenko_2016_aa},
$^c$\cite{gaia-collaboration_2018_aa},
$^d$\cite{harper_2017_aa},
$^e$\cite{schiller_2008_aa},
$^f$\cite{tetzlaff_2011_aa},
$^g$\cite{howarth_2001_aa},
$^h$\cite{carpenter_1999_aa},
$^i$\cite{cvetkovic_2009_aa},
$^j$\cite{shultz_2014_aa},
$^k$\cite{montarges_2018_aa},
$^l$\cite{ohnaka_2013_aa},
$^m$\cite{neilson_2011_aa},
$^n$\cite{van-belle_2009_aa,malagnini_2000_aa},
$^o$\cite{north_2007_aa},
$^p$\cite{lee_2014_aa},
$^q$\cite{baines_2018_aa},
$^r$\cite{reimers_1989_aa},
$^s$\cite{tokovinin_1997_aa}
}
\end{deluxetable*}

\begin{deluxetable*}{ccccc}[!htb]
\renewcommand{\arraystretch}{1.05}
\tablecolumns{8}
\tablecaption{Minimum Angular Separation Between Pre-supernova Candidates.}
\label{tab:angularsep}
\tablehead{
\colhead{N} & \colhead{Catalog/Common} &  \colhead{Min. Ang.} & \colhead{Nearest Neighbor} & \colhead{Nearest Neighbor} \\
            & Name                     &  Separation (degree) & Name                       & Number 
}
\startdata
1  & HD 116658/Spica & 39.66 & HD 129056/$\alpha$ Lupi & 3 \\
2  & HD 149757/$\zeta$ Ophiuchi &  15.97 & HD 148478/Antares & 5\\
3  & HD 129056/$\alpha$ Lupi & 29.73 & HD 148478/Antares & 5\\
4  & HD 78647/$\lambda$ Velorum & 1.73 & HD 80108/HR 3692 & 27 \\
5  & HD 148478/Antares & 15.97 & HD 149757/$\zeta$ Ophiuchi & 2\\
6  & HD 206778/$\epsilon$ Pegasi & 13.08 & HD 207089/12 Pegasi & 19 \\
7  & HD 39801/Betelgeuse & 11.59 &  S Mono A/B & 12/13\\
8  & HD 89338/q Car & 3.30 & HD 93070/w Car & 14 \\
9  & HD 210745/$\zeta$ Cephei & 5.69 & HD 208816/VV Cephei & 22 \\
10 & HD 34085/Rigel & 18.60 & HD 39801/Betelgeuse & 7\\
11 & HD 200905/$\zeta$ Cygni & 4.39 & HD 102098/Deneb & 31 \\
12 & HD 47839/S Mono A & 11.60 & HD 39801/Betelgeuse & 7\\
13 & HD 47839/S Mono B & 11.60 & HD 39801/Betelgeuse & 7 \\
14 & HD 93070/w Car & 3.30 & HD 89338/q Car & 8\\
15 & HD 68553/NS Puppis & 7.72 & HD 68273/$\gamma^2$ Velorum & 17 \\
16 & HD 36389/119 Tauri & 12.50 & HD 39801/Betelgeuse & 7\\
17 & HD 68273/$\gamma^2$ Velorum & 7.72 & HD 68553/NS Puppis & 15 \\
18 & HD 50877/$o^1$ Canis Majoris & 4.12 & HD 52877/$\sigma$ Canis Majoris & 21 \\
19 & HD 207089/12 Pegasi & 13.08 & HD 206778/$\epsilon$ Pegasi & 6 \\
20 & HD 213310/5 Lacertae & 4.88 & HD 216946/V424 Lacertae & 25 \\ 
21 & HD 52877/$\sigma$ Canis Majoris & 4.12 & HD 50877/$o^1$ Canis Majoris & 18\\
22 & HD 208816/VV Cephei & 5.69 & HD 210745/$\zeta$ Cephei & 9 \\ 
23 & HD 196725/$\theta$ Delphini & 16.39 & HD 206778/$\epsilon$ Pegasi & 6\\ 
24 & HD 203338/V381 Cephei & 6.72 & HD 208816/VV Cephei & 22\\
25 & HD 216946/V424 Lacertae & 4.88 & HD 213310/5 Lacertae & 20\\
26 & HD 17958/HR 861 & 23.49 & HD 219978/V809 Cassiopeia & 29\\
27 & HD 80108/HR 3692 & 1.73 & HD 78647/$\lambda$ Velorum & 4 \\
28 & HD 56577/145 Canis Majoris & 5.22 & HD 50877/$o^1$ Canis Majoris & 18 \\
29 & HD 219978/V809 Cassiopeia & 9.33 & HD 208816/VV Cephei & 22 \\
30 & HD 205349/HR 8248 & 5.38 & HD 200905/$\zeta$ Cygni & 11 \\
31 & HD 102098/Deneb & 4.39 & HD 200905/$\zeta$ Cygni & 11 \\
\enddata
\end{deluxetable*}

\clearpage

\bibliography{refs}
\bibliographystyle{aasjournal}

\end{document}